\documentclass[11pt,preprintnumbers,superscriptaddress,amsmath,amssymb,nofootinbib]{revtex4-1}
\usepackage{graphicx}
\usepackage{dcolumn}
\usepackage{bm}
\usepackage{amssymb}
\usepackage{amsmath}
\usepackage{epsfig}    
\usepackage{color}
\usepackage{slashed}
\usepackage{hyperref}
\usepackage{multirow,array}

\usepackage{physics}
\usepackage{listings}
\usepackage{cprotect}

\allowdisplaybreaks[1]

\begin{document} 

% \title{H-COUP Version 3:\\ A program for one-loop corrected Higgs boson decays\\
% in non-minimal Higgs sectors}

\title{H-COUP Version 3:\\ A program for one-loop corrected decays of any Higgs bosons\\ in non-minimal Higgs models}

%\date{\today}

\preprint{KEK-TH-2578, OU-HET 1201, NU-EHET 002, TU-1214}

\author{Masashi Aiko}
\email{maiko@post.kek.jp}
\affiliation{KEK Theory Center, IPNS, KEK, Tsukuba, Ibaraki 305–0801, Japan}

\author{Shinya Kanemura}
\email{kanemu@het.phys.sci.osaka-u.ac.jp}
\affiliation{Department of Physics, Osaka University, Toyonaka, Osaka 560-0043, Japan}
\author{Mariko Kikuchi}
\email{kikuchi.mariko13@nihon-u.ac.jp}
\affiliation{College of Engineering, Nihon University, Koriyama, Fukushima 963-8642, Japan}

\author{Kodai Sakurai}
\email{kodai.sakurai.e3@tohoku.ac.jp}
\affiliation{Institute of Theoretical Physics, Faculty of Physics, University of Warsaw, ul. Pasteura 5, PL-02-093 Warsaw, Poland}
\affiliation{Department of Physics, Tohoku University, Sendai, Miyagi 980-8578, Japan}

\author{Kei Yagyu}
\email{yagyu@het.phys.sci.osaka-u.ac.jp}
\affiliation{Department of Physics, Osaka University, Toyonaka, Osaka 560-0043, Japan}

\begin{abstract}

The {\tt H-COUP} program is provided as a package of Fortran codes, which can compute observables related to Higgs bosons including radiative corrections
in various extended Higgs sectors.
We give a manual for the latest version of {\tt H-COUP} (\verb|H-COUP_3.0|), in which 
decay rates and branching ratios of all the Higgs bosons can be calculated at one-loop level in EW and Higgs interactions with QCD corrections 
in the Higgs singlet model, four types of the two Higgs doublet model with a softly-broken $Z_2$ symmetry, and the inert doublet model. 
The previous version (\verb|H-COUP_2.0|) can evaluate those only for the standard model like Higgs boson with the mass of 125 GeV ($h$). 
In \verb|H-COUP_3.0|, renormalized quantities are computed based on the gauge independent on-shell renormalization scheme. 
The source code of \verb|H-COUP_3.0| can be downloaded via the following link: \url{http://www-het.phys.sci.osaka-u.ac.jp/~hcoup}.
By using \verb|H-COUP_3.0|, we can compare the precise measurements of the properties of $h$ and direct searches for additional Higgs bosons 
with their predictions at one-loop level, by which we can reconstruct the structure of the Higgs sector. 
%global vision
%
\end{abstract}
\maketitle
%\tableofcontents

\section{Introduction} \label{sec: introduction}

The LHC experiments have clarified the existence of the 125 GeV Higgs boson and its properties 
to be consistent with those of the Higgs boson in the Standard Model (SM) within the theoretical and experimental uncertainties~\cite{ATLAS:2022vkf, CMS:2022dwd}.
However, the nature of electroweak symmetry breaking is still one of the most important questions in particle physics. 
In fact, no fundamental principle has been proved
for determining the structure of the Higgs sector.
Indeed, the SM merely assumes the minimal form composed of an isospin scalar doublet field. 
Thus, there is no compelling reason to stay at the minimal form, and are various possibilities for extended structures of the Higgs sector. 
On the other hand, the SM obviously cannot explain tiny neutrino masses, existence of dark matter and baryon asymmetry of the Universe,
so that new physics must exist. 
The question is then what is the scale of new physics, which could be much higher than the electroweak scale, e.g., in the conventional seesaw scenario or could be EW/TeV scales, e.g., in the electroweak baryogenesis. 
If the latter is realized, its experimental verification is expected.  
In particular, it is quite natural to think that the Higgs sector is modified from the minimal form. 
In fact, extended Higgs sectors are often introduced in  new physics at TeV scale, e.g., models with an extra isospin singlet, doublet and triplet field and so on, and their property strongly depends on new physics scenario.   
Therefore, unveiling the true structure of the Higgs sector is a key to open the door to new physics. 

%it is quite challenging to directly test such a scenario. 
%(scale is unknown High scale or Low scale, if low scale appear then 
%it is natural to consider the Higgs sector then we can test it by 
%experiments. )
%
%In such a new physics scenario, the Higgs sector is often extended from %the minimal form

%in the SM, the wine-bottle type potential is assumed to realize the electrothe Higgs sector  
%does not have a mechanism to realize , 
%and the introduction of the scalar field causes unnatural fine-tuning of the Higgs boson mass. 
%It is thus quite natural to consider a new paradigm behind the Higgs sector such as supersymmetry, dynamical symmetry breaking and/or extra dimensions. 
%Although none of them has not been probed yet, a clue of these new paradigms can be found through the Higgs sector, because of predictions of its extended structure at the electroweak scale. 
%

The structure of the Higgs sector can be determined by investigating various extended Higgs sectors comprehensively. % as a bottom-up approach. 
There are basically two ways for such an investigation, i.e., precise measurements for properties of the discovered Higgs boson ($h$) and direct searches for extra Higgs bosons. 

The former is absolutely important, because the precise measurement of $h$ is what we can definitely perform at future collider experiments such as the High-Luminosity LHC (HL-LHC)~\cite{Apollinari:2017lan} and 
lepton colliders, e.g., the International Linear Collider (ILC)~\cite{Baer:2013cma,Fujii:2017vwa,Asai:2017pwp,Fujii:2019zll}, the Circular Electron-Positron Collider (CEPC)~\cite{CEPC-SPPCStudyGroup:2015csa} and $e^+e^-$ collisions of the Future Circular Collider (FCC-ee)~\cite{Gomez-Ceballos:2013zzn}. 
For example, the Higgs boson couplings are expected to be measured typically  with a few percent level at HL-LHC~\cite{deBlas:2019rxi} and a few permille level at $e^+e^-$ colliders~\cite{deBlas:2019rxi}. 
Thus, predictions of the Higgs boson couplings, decay rates and branching ratios in extended Higgs sectors must be evaluated at quantum levels in order to compare such precise measurements. 
%%%
Once deviations in the observables of $h$ from SM predictions are found, 
we can extract the scale of new physics from the size of deviations, which has been known as  a ``new no-loose theorem''~\cite{Kanemura:2021fvp} . 
Moreover, we can fingerprint the Higgs sector, i.e., extracting the properties such as 
the representation and the number of Higgs fields by measuring the pattern of deviations. 

Detections of additional Higgs bosons are direct evidence of an extended Higgs sector.
%, because their appearance is 
%the common feature of extended structures.
Although any clear signatures have not been observed yet, there are still possibilities of the existence of extra Higgs bosons
at the electroweak scale, which is well motivated by various new physics scenarios, e.g., the radiative seesaw mechanism, the Higgs-portal dark matter and the electroweak baryogenesis. 
It has been known that the decay property of extra Higgs boson strongly depends on the ``alignmentness'' of the Higgs sector, i.e., 
how the property of $h$ is close to the one in the SM Higgs boson. 
In the nearly alignment case, extra Higgs bosons tend to dominantly decay into a bosonic final state, e.g., $H \to W^+W^-/ZZ/hh$, $A \to Zh$ and $H^\pm \to W^\pm h$ in the Two Higgs Doublet Models (THDMs), 
in which severe lower limits on the mass of extra Higgs bosons have already been taken at the LHC~\cite{Aiko:2020ksl,Cheung:2022ndq,Arhrib:2016wpw}. 
This also means that a lower limit and an upper limit on the mass of extra Higgs bosons 
can simultaneously be taken by the synergy between the direct searches at the HL-LHC and the precise measurements of $h$, 
and thus a large portion of the parameter space can be explored~\cite{Aiko:2020ksl}. 
The important thing here is that loop effects of extra Higgs bosons
can change the alignmentness and the decay rates of extra Higgs bosons at the same time. 
In particular, such modifications can be significant when non-decoupling effects of extra Higgs bosons are realized, which 
can appear in new physics scenarios, e.g., with first order electroweak phase transitions~\cite{Kanemura:2002vm,Kanemura:2004ch,Kanemura:2004mg}. 
Therefore, including the effect of radiative corrections is important not only for the precise measurement of $h$, but also for the direct searches for extra Higgs bosons. 

In order to realize the fingerprinting, to establish the new no-loose theorem and to perform the direct searches in a more precise way, 
%precise calculations of the Higgs sector and to make the ``synergy analysis'', 
%i.e., plain) more robust, % (fingerprinting and no-loose theorem)
we have developed the \verb|H-COUP| program which is provided as a package of Fortran codes to evaluate 
Higgs boson related observables including radiative corrections in various extended Higgs sectors. 
So far, we have published the \verb|H-COUP| version 1 (\verb|H-COUP_1|)~\cite{Kanemura:2017gbi} and version 2 (\verb|H-COUP_2|)~\cite{Kanemura:2019slf}, 
which can provide one-loop corrected couplings, decay rates and branching ratios of $h$ in the THDMs with a softly-broken $Z_2$ symmetry to avoid flavor changing neutral currents at tree level, the Higgs Singlet Model (HSM)
and the Inert Doublet Model (IDM) based on the gauge independent on-shell renormalization scheme~\cite{Kanemura:2017wtm}. 

In this paper, we extend the \verb|H-COUP| program to version 3 (\verb|H-COUP_3.0|)
and give its manual, where  
the decay rates and the branching ratios of {\it all} the Higgs bosons can be evaluated at one-loop level in the extended Higgs sectors shown above. 
%%% 
We have implemented one-loop corrected decay rates of extra Higgs bosons in \verb|H-COUP_3.0|, which have been computed in the series of our papers 
\cite{Kanemura:2022ldq,Aiko:2022gmz,Aiko:2021can}. % for $H$, $A$ and $H^\pm$, respectively, in the THDMs. 
%In order to develop , we have computed the radiative corrections to the decay rates of extra Higgs bosons 
%
In \verb|H-COUP_3.0|, we take the same on-shell scheme for the scalar two-point functions as the previous versions, while the renormalization for the tadpole
is performed by either the standard tadpole scheme~\cite{Hollik:1988ii, Denner:1991kt}
 or the alternative tadpole scheme~\cite{Fleischer:1980ub} (in the case of the THDMs, there are four choices because of the renormalization method of the other parameter).
%$\overline{\text{MS}}$ parameter).
The users can choose these renormalization schemes. 
%from several choices such as 
%
%In \ver, we argue the difference in renormalizations of tadpole, 
%so that we give users a choice of renormalization schemes of tadpole for the $\phi\to \phi'\phi''$′ decay in the HSM and the THDM. 
%
%We describe structures of the loop corrected decay rates for a set of decay types for scalar bosons; 
%i.e. $\phi\to \phi'\phi''$, $\phi\to V\phi'$ $\phi\to VV'$ and $\phi\to f\bar{f'}$, and decay rates of loop induced processes; 
%i.e. $\phi\to V\gamma$, $\phi\to gg$ and $\phi\to W^\pm Z$.
%
There are also important works relevant to such radiative corrections in the THDMs~\cite{Dao:2019nxi,Krause:2019qwe,Oliveira:2001vw,Akeroyd:1998uw,Akeroyd:2000xa,Krause:2016oke,Krause:2016xku,Santos:1996hs} and in the minimal supersymmetric SM~\cite{Barger:1991ed,Chankowski:1992es,Osland:1998hv,Philippov:2006th,Williams:2011bu,Williams:2007dc}.

We note that similar program tools have been developed by the different groups, e.g., {\tt 2HDECAY}~\cite{Krause:2018wmo}, {\tt Prophecy4f}~\cite{Denner:2019fcr}, {\tt ewN2HDECAY}~\cite{Krause:2019oar}  and {\tt EWsHDECAY}~\cite{Egle:2023pbm}\footnote{Recently, {\tt FlexibleDecay}~\cite{Athron:2021kve} appeared, in which decays of Higgs bosons can be computed in the $\overline{\rm MS}$ scheme with higher order electroweak and QCD corrections in various models beyond the SM. }.
A remarkable feature of {\tt H-COUP} is that it can systematically compute 
both the properties of $h$ and extra Higgs bosons in various extended Higgs sectors under a fixed renormalization scheme. %%% More emphasize! 

This paper is organized as follows. In Sec.~\ref{sec: model}, 
we briefly review the extended Higgs models which are implemented in \verb|H-COUP_3.0| to define the notation. 
In Sec.~\ref{sec: reno}, we discuss the renormalized vertex functions and decay rates. We also explain our renormalization scheme. 
We then describe the structure of \verb|H-COUP_3.0| and how to install and run the program in Sec.~\ref{sec: structure}.
In Sec.~\ref{sec:Discussion}, we show examples of the numerical evaluation. 
Summary is given in Sec.~\ref{sec:summary}.

%%%%%%%%%%%%%%%%%%%%%%%%%%%%%%%%%%%%%%%%%%%%%%%%%%%%%%%
\section{Models and constraints} \label{sec: model}

We define three models with the extended Higgs sector, i.e, the HSM, the THDMs and the IDM which are implemented in \verb|H-COUP_3.0|. 
In Table~\ref{tab:models}, we show the mass eigenstates of scalar fields, the input parameters and the constraints in each model.
In the following, we provide additional explanations of Table~\ref{tab:models} in each model, see the references given in this table for more details 
of the model description.
%~\footnote{Although only free parameters beyond the SM are shown in Table I, there are also the mass of the discovered Higgs boson ($m_h\simeq125$ GeV) and the vacuum expectation value (VEV) ($v\simeq 246$ GeV) in addition to the new physics parameters.}. 

%In Table~\ref{tab:models}, mass eigenstates of scalar fields, free parameters used as input parameters in \verb|H-COUP| and references describing the details of the models are shown.

Throughout the paper, we denote the mass eigenstates of scalar fields as: 
\begin{align}
 h \space{}&:\space{}\textrm{the discovered CP-even Higgs boson with the mass 125 GeV, } \notag\\
 H \space{}&:\space{}\textrm{another CP-even Higgs boson, } \notag\\
 A \space{}&:\space{}\textrm{a CP-odd Higgs boson, }\\
 H^\pm \space{}&:\space{}\textrm{a pair of singly charged Higgs bosons}.  \notag
\end{align} 
%where these symbols are used uniformly in all the models. 
%In addition, any scalar particle is represented by $\phi$. 

\begin{table}[t]
\begin{center}
{\renewcommand\arraystretch{1.2}
\begin{tabular}{c||c|c|c|c|c|c|c}\hline\hline
  \multirow{2}{*}{Models} & \;\multirow{2}{*}{Mass eigenstates}\; & \multirow{2}{*}{Input parameters} & \multicolumn{5}{c}{\;References for constraints\;}  \\
\cline{4-8}  
  && & \;(a), (b)\; & \;(c)\; & \;(d)\; & \;(e)\;  &\;(f)\;\\
  \hline\hline
  HSM~\cite{Kanemura:2015fra,Kanemura:2016lkz} & $h, \; H$ & $ \;m_H,\; \alpha,\; \lambda_S^{},\; \lambda_{\Phi S},\; \mu_S^{}\;$ 
  & \cite{Pruna:2013bma,Fuyuto:2014yia,Robens:2015gla} %Vacuum stability (HSM)
  &\cite{Cynolter:2004cq} % Tree level unitarity(HSM)
  &\cite{Chen:2014ask,Kanemura:2016lkz}
   % Tree vacuum (HSM)
  & \cite{Kanemura:2016lkz,Gonderinger:2009jp}
  % Triviality (HSM)
  &\cite{Lopez-Val:2014jva} % EW ST (HSM)
  \\\hline
 \multirow{1}{*}{THDMs~\cite{Kanemura:2004mg, Kanemura:2014dja, Kanemura:2015mxa}}&
  \multirow{2}{*}{$h,\; H,\; A,\; H^\pm $} &  $m_H^{},\; m_A^{},\; m_{H^\pm}^{},\;M^2,\;$  
  &\multirow{2}{*}{\cite{Deshpande:1977rw,Klimenko:1984qx,Sher:1988mj,Nie:1998yn,Kanemura:1999xf}}
  %Vacuum stability (THDM)
  & \multirow{2}{*}{\cite{Kanemura:1993hm,Akeroyd:2000wc,Ginzburg:2005dt,Kanemura:2015ska}}
  % Tree level unitarity(THDM)
  &\multirow{2}{*}{\cite{Branchina:2018qlf}}
  % Tree vacuum (THDM)
  &\multirow{2}{*}{\cite{Inoue:1982ej}}% Triviality (THDM)
  &\multirow{2}{*}{\cite{Toussaint:1978zm,Bertolini:1985ia,Peskin:2001rw,Grimus:2008nb,Kanemura:2011sj}}
  % EW ST (THDM)
  \\
  (Type-I, II, X, Y)\;&&\;$ \sin(\beta-\alpha),\; \tan\beta$\; &&&&& \\\hline
  IDM~\cite{Kanemura:2016sos} & $h,\; H,\; A,\; H^\pm $ & $ m_H^{},\; m_A^{},\; m_{H^\pm}^{},\; \mu_2^{2},\; \lambda_2^{}$ 
  &\cite{Deshpande:1977rw,Klimenko:1984qx,Sher:1988mj,Nie:1998yn,Kanemura:1999xf}
  %Vacuum stability (IDM)
  &\cite{Kanemura:1993hm,Akeroyd:2000wc,Ginzburg:2005dt,Kanemura:2015ska}
  % Tree level unitarity(IDM)
  &\cite{Ginzburg:2010wa}
   % Tree vacuum (IDM)
  & \cite{Goudelis:2013uca} % Triviality (IDM)
  &\cite{Toussaint:1978zm,Bertolini:1985ia,Peskin:2001rw,Grimus:2008nb,Kanemura:2011sj}
  % EW ST (IDM)
  \\\hline\hline
\end{tabular}}
\caption{Mass eigenstates of scalar fields, input free parameters and references in the HSM, the THDMs and the IDM. Although only free parameters beyond the SM are given, there are two SM parameters in the Higgs potential; i.e. the mass of the discovered Higgs boson ($m_h\simeq125$ GeV) and the vacuum expectation value (VEV) ($v\simeq 246$ GeV), in each model. The numbers in ``References of constraints'' indicate kinds of constraints as; (a) Vacuum stability at tree level, (b) Vacuum stability (RGE improved), (c) Tree-level unitarity, (d) True vacuum, (e) Triviality, (f) Electroweak $S$ and $T$ parameters. 
%\textcolor{red}{[Have to correct placement of literature !]}
}
\label{tab:models}
\end{center}
\end{table}

In the HSM, the Higgs sector is composed of the SM Higgs field $\Phi$, i.e., the isospin doublet
Higgs field with hypercharge $Y = 1/2$, and an isospin singlet scalar field $S$ with $Y = 0$. 
In this model, no additional symmetry beyond the SM is imposed~\cite{Chen:2014ask}. 
There are two mass eigenstates and 
five free input parameters as shown in Table~\ref{tab:models},  
where $m_\phi$ indicates a mass of a scalar boson $\phi$,   
and $\alpha$ is the mixing angle between two CP-even Higgs bosons with its domain to be defined as $-\pi/2 \leq \alpha \leq \pi/2$. 
The other three parameters $\lambda_S$, $\lambda_{\Phi S}$
and $\mu_S$ are the coefficients of the $S^4$, $|\Phi|^2S^2$ 
and $S^3$ terms in the Higgs potential, respectively.

The THDMs contain two isospin doublet Higgs fields $\Phi_1$ and $\Phi_2$ with $Y = 1/2$. 
In \verb|H-COUP|, we impose a softly-broken $Z_2$ symmetry~\cite{Glashow:1976nt} to avoid flavor changing neutral currents at tree level, in which 
$\Phi_1$ and $\Phi_2$ are assigned to be $Z_2$-even and $Z_2$-odd, respectively.
Depending on the charge assignments of right-handed fermions, 
four types of Yukawa interactions appear, which are so-called 
Type-I, Type-II, Type-X and Type-Y~\cite{Barger:1989fj,Grossman:1994jb,Akeroyd:1996he,Aoki:2009ha}. %
There are five mass eigenstates and 
six input parameters described in Table~\ref{tab:models}, where
$\alpha$ is the mixing angles between two CP-even Higgs bosons and $\tan\beta$ is the ratio of the VEVs $\tan\beta = v_2/v_1$~\footnote{In addition, we have to input the sign of $\cos(\beta - \alpha)$ because values of $\alpha$ and $\beta$ are not directly input.} 
We define $\sin(\beta-\alpha)\geq 0$ and $\tan\beta >0$.   
A dimensionful parameter $M^2$ describes the soft breaking scale of the $Z_2$ symmetry. 

In the IDM, the Higgs sector also consists of two isospin doublet Higgs fields $\Phi$ and $\eta$
with $Y = 1/2$, 
so that five mass eigenstates appear as those in the THDMs. 
Unlike the THDMs, an unbroken $Z_2$ symmetry~\cite{Deshpande:1977rw,Barbieri:2006dq} is imposed instead of the softly-broken one, in which only $\eta$ is assigned to be $Z_2$-odd and all the other fields are assigned to be $Z_2$-even.  
Thanks to the unbroken $Z_2$ symmetry, the lightest additional neutral scalar boson ($H$ or $A$) can be a candidate for dark matter. 
There are five free parameters shown in Table~\ref{tab:models}, where  
$\lambda_2$ and $\mu_2^2$ are the coefficients of the quartic and the quadratic terms of $\eta$, respectively.

In each model, the following constraints can be imposed in \verb|H-COUP_3.0| as in the previous versions: 

\begin{enumerate}
\renewcommand{\labelenumi}{(\alph{enumi})}
    \item Vacuum stability at tree level~\cite{Deshpande:1977rw}
    \item Vacuum stability (Renormalization group equations (RGE) improved with the cutoff)
    \item Tree-level unitarity~\cite{Lee:1977eg}
    \item True vacuum~\cite{Espinosa:2011ax,Branchina:2018qlf}
    \item Triviality (with the cutoff scale)~\cite{Inoue:1982ej}.  
    \item Electroweak $S$ and $T$ parameters~\cite{Peskin:1990zt,Peskin:1991sw}
\end{enumerate}
%which are the same constraints as those of \verb|H-COUP_1.0| and  \verb|_2.0|. 
In Table~\ref{tab:models}, we list the references for these constraints in each model, see also the manual of \verb|H-COUP_1.0|~\cite{Kanemura:2017gbi} for 
details. 
%We also give descriptions of above the six constraints in collectively.

%%%%%%%%%%%%%%%%%%%%%%%%%%%%
\section{Renormalized vertices and decay rates} \label{sec: reno}

% In this section, we discuss radiative corrections to the decay rate of a scalar boson $\phi$ in a model-independent way. 
In this section, we discuss radiative corrections to the decay rate of a scalar boson $\phi$ after introducing renormalization schemes implemented in \verb|H-COUP_3.0|. 
While the discussions of the renormalization schemes are model-dependent, the formulation of the decay rate is carried out in a model-independent way.  \par\vspace{2ex}

\subsection{ Renormalization schemes}

The renormalization of Higgs potential parameters is performed by making use of on-shell conditions for two-point functions of the scalar states. 
With the on-shell conditions, masses, mixing angles, and wave functions renormalization factors for the scalar states are defined as on-shell parameters.
In Ref.~\cite{Yamada:2001px}, it has been pointed out that the on-shell renormalization of scalar mixing angles give rise to gauge-dependent amplitudes.
In order to remove such gauge dependence, we make use of the pinch technique~\cite{Papavassiliou:1994pr}, in which appropriate pinch terms are added for the counterterms of the mixing angles. 

The UV divergence of the one-point functions of Higgs fields should also be taken away in the renormalized Higgs potential. 
For the renormalization of tadpoles, there are two different renormalization schemes, which are called the standard tadpole scheme~\cite{Hollik:1988ii, Denner:1991kt} and the alternative tadpole scheme~\cite{Fleischer:1980ub} 
(also see a new scheme for tadpole renormalization in recent works~\cite{Dittmaier:2022maf, Dittmaier:2022ivi}). 
Let us briefly describe how the renormalization of tadpoles is performed in these two schemes. 
In the standard tadpole scheme, {\it renormalized} tadpoles are set to zero. 
One then has the tadpole counterterms, which eliminate the UV divergence for the one-point functions by renormalization conditions for tadpoles, i.e., $\hat{\Gamma}_h=\Gamma_h^{\rm 1PI}+\delta T_h=0$ for the SM-like Higgs boson $h$. 
In contrast, in the alternative tadpole scheme, {\it unrenormalized} tadpoles are set to zero, which means that the tadpole counterterm parameters are not introduced in a theory.
Instead, there is a degree of freedom to shift bare Higgs fields, $\phi_B\to \phi_B+\Delta v$. 
This shift introduces an additional tadpole term, and one can choose the constant $\Delta v$ in a way that the tadpole terms vanish at loop level, i.e., $\Gamma_h^{\rm 1PI}+m_h^2\Delta v=0$ for the SM-like Higgs boson $h$. 
The shift affects all terms containing the Higgs bare field $\phi_B$. 
Because of this, one eventually should include tadpole-inserted diagrams in renormalized self-energies and renormalized vertex functions. 

In all the models implemented in \verb|H-COUP_3.0|, the number of input parameters is greater than that determined by the on-shell conditions for the scalar two-point functions.
Thus, the other remaining parameters should be renormalized in different ways. 
As briefly discussed, such a parameter is determined by the $\overline{\rm MS}$ renormalization. 
In the following, by specifying a model, we discuss how each input parameter is renormalized 
and introduce renormalization schemes implemented in \verb|H-COUP_3.0|. 
The renormalization of model parameters is summarized in Table~\ref{tab:CTs}. 

\begin{table}[t]
\begin{center}
{\renewcommand\arraystretch{1.2}
\begin{tabular}{c||c|c|c}\hline\hline
  Models & \;On-shell renormalization\; & $\overline{\rm MS}$ renormalization & Refs.  \\\hline
  HSM & $\;\delta m^2_h,\delta m^2_H,\; \delta \alpha,\delta Z_\phi, \delta C_\phi\;$ & $  \delta\lambda_{\Phi S},\; \delta \mu_S^{}\;$ & \cite{Kanemura:2016lkz} \\\hline
  THDMs & $\delta m_h^{2},\delta m_H^{2},\; \delta m_A^{2},\; \delta m_{H^\pm}^{2},\; \delta\alpha,\; \delta\beta,\;\delta Z_\phi, \delta C_\phi\;$ &  $\delta M^2$ (or $\delta m_{12}^2$) & \cite{Kanemura:2015mxa} \\\hline
  IDM & $ \delta m_h^{2},\;\delta m_H^{2},\; \delta m_A^{2},\; \delta m_{H^\pm}^{2},\; \delta Z_\phi $ & $ \; \delta\mu_2^{2}$ & \cite{Kanemura:2016sos} \\\hline\hline
\end{tabular}}
\caption{Counterterms used in the computations of decays of scalar bosons in the HSM, the THDMs, and the IDM. 
In the wave faction factors $\delta Z_{\phi}$ and $\delta C_{\phi}$, the subscript ``${\phi}$ '' denotes any of Higgs bosons listed in Table.~\ref{tab:models} for each model. If one chooses the standard tadpole scheme, the counterterms for tadpoles should be also added. 
Detailed discussions about the renormalization of these counterterms can be found in the quoted references.
The gauge invariant renormalization of the mixing angles $\alpha$ and $\beta$ can be found in Ref.~\cite{Kanemura:2017wtm}. }
\label{tab:CTs}
\end{center}
\end{table}

\noindent{\underline{Higgs singlet model}}

In the HSM, there are two mass parameters (i.e., $m_h$ and $m_H$), one mixing angle and four parameters of wave function renormalizations. 
They are renormalized by on-shell conditions for two-point functions of $h$ and $H$, and the other parameters $\lambda_{\Phi S}$ and $\mu_S$ are determined by the $\overline{\rm MS}$ renormalization. 
 % The counter term of singlet VEV $v_s$ is taken to be zero in in \verb|H-COUP_3.0|. 
 % Although it appears in the 1-loop corrections to $H\to hh$ and the renormalized $hhh$ vertex function as long as the bare parameter v_{s,B} is nonzero
 Since there are two different ways to renormalize tadpoles as discussed above, 
in \verb|H-COUP_3.0|, a user can choose the following renormalization schemes for the tadpoles in the HSM: 
\begin{align} 
\mbox{ [KOSY]:} &\mbox{ Standard tadpole scheme }\nonumber \\
\mbox{ [PT]:} &\mbox{ Alternative tadpole scheme} \nonumber
 \end{align}
The KOSY and the PT scheme schemes are respectively based on Ref.~\cite{Kanemura:2004mg} and Ref.~\cite{Kanemura:2017wtm}. 
We here emphasize that the difference between these two renormalization schemes is how the tadpoles are renormalized. 
For all the above schemes, the scalar mixing angle $\alpha$ is commonly renormalized by the on-shell scheme with the pinch terms. 
%This parameter is related to 

\noindent{\underline{Two Higgs doublet models}}

In the THDMs, four mass parameters, two mixing angles, and twelve wave functions are determined by the on-shell renormalization. 
The reaming parameter $M^2$ is determined by the $\overline{\rm MS}$ renormalization.
This parameter is related to the softly-broken parameter of the $Z_2$ symmetry $m_{12}^2$ ($V\supset (m_{12}^2\Phi_1^\dagger \Phi_2 +{\rm h.c.})$ ) by $M^2=m_{12}^2/{(\cos\beta \sin\beta)}$.  
Instead of $M^2$, one can also renormalize $m_{12}^2$ as a $\overline{\rm MS}$ parameter. 
% some of other parameters are defined by $\overline{\rm MS}$ scheme, i.e., $M_{12}$ for THDMs, $\lambda_{\Phi S}$, $\mu_S$ for HSM, and $\mu_2$ for IDM. 
% \footnote{renormalization of $\lambda_s$ in HSM and $\lambda_2$ in IDM is not required in the computation of one-loop corrections to Higgs boson decays. }
%There are several choices for renormalization of tadpoles and $\overline{\rm MS}$ parameters.  
% \red{[memo:Explantion of standard tadpole and alternative tadpole,]}
% For the renormalization of tadpoles, there are two different ways which discussed in earlier works, standard tadpole scheme~\cite{} and alternative tadpole scheme~\cite{}. 
% In addition, for the $\overline{\rm MS}$  parameters, one has choices which parameters one renormalizes. 
% For example in THDMs, one can renormalize $m_{12}^2$, instead of $M^2$, where $m_{12}^2$ is the softly broken parameter of $Z_2$ symmetry ($V\supset (m_{12}^2\phi_1 \phi_2^\dagger +{\rm h.c.})$ ) and is related with $M$ by $M^2=m_{12}^2/{(c_\beta s_\beta)}$. 
In \verb|H-COUP_3.0|, a user can choose the renormalization scheme for the tadpoles and the $\overline{\rm MS}$ parameter in the THDMs. 
Hence, the following four different renormalization schemes are implemented:
\begin{align}
  \mbox{[KOSY1]:} &\mbox{ Standard tadpole scheme, }\delta M^2  \nonumber\\
   \mbox{[PT1]:} &\mbox{ Alternative tadpole scheme, }\delta M^2  \nonumber\\
   \mbox{[KOSY2]:}&\mbox{ Standard tadpole scheme, }\delta m_{12}^2 \nonumber\\
   \mbox{[PT2]:} &\mbox{ Alternative tadpole scheme, }\delta m_{12}^2 \nonumber
\end{align}
 The KOSY1 and KOSY2 schemes are based on Ref.~\cite{Kanemura:2004mg}, while the PT1 and PT2 schemes are based on Ref.~\cite{Kanemura:2017wtm}.
In all these renormalization schemes, the pinch technique is applied to the renormalization of the mixing angles $\alpha$ and $\beta$. 
%all the on-shell parameters are renormalized in the same way. 

\noindent{\underline{Inert doublet model}}

In the IDM, four mass parameters and four wave functions are determined by the on-shell renormalization. 
The invariant mass parameter $\mu_2^2$ for the inert scalars is renormalized as a $\overline{\rm {MS}}$ parameter. 
In Ref.~\cite{Kanemura:2017wtm}, it has been shown that for renormalized vertex functions of $h$, the scheme difference between the standard tadpole scheme and the alternative tadpole scheme does not appear in the SM. 
This is also the case in the IDM, because the counterterms for the renormalized vertex functions are expressed in the same form as those in the SM. 
Hence, for the IDM, we only implement the KOSY scheme based on Ref.~\cite{Kanemura:2016sos} in \verb|H-COUP_3.0|.

\subsection{ Renormalized vertex functions}

\begin{figure}
    \centering
    \includegraphics[width=60mm]{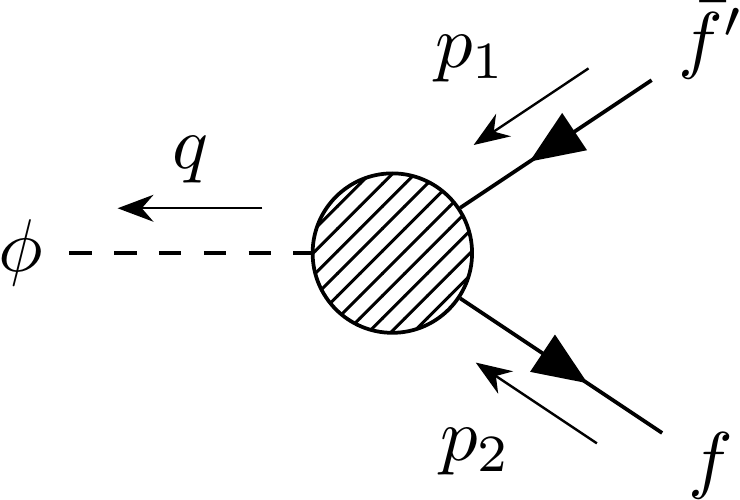}
    \includegraphics[width=60mm]{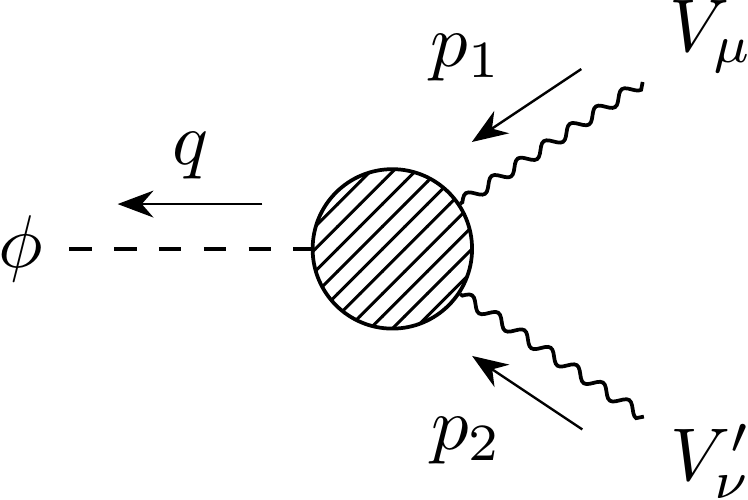}\\ \vspace{5mm}
    \includegraphics[width=60mm]{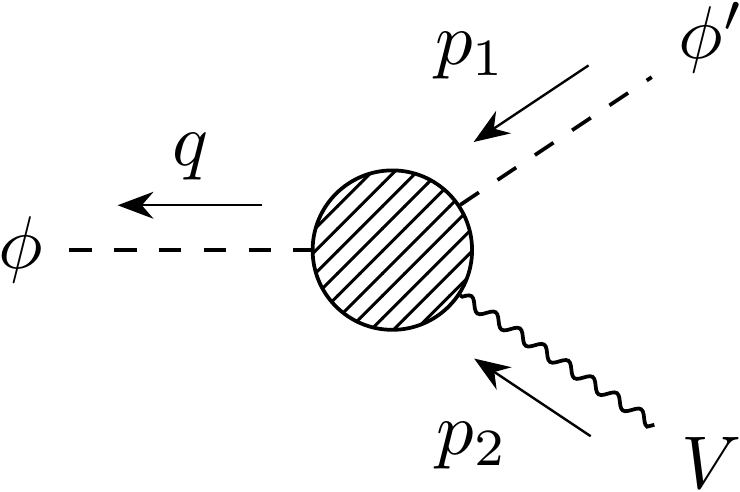}
    \includegraphics[width=60mm]{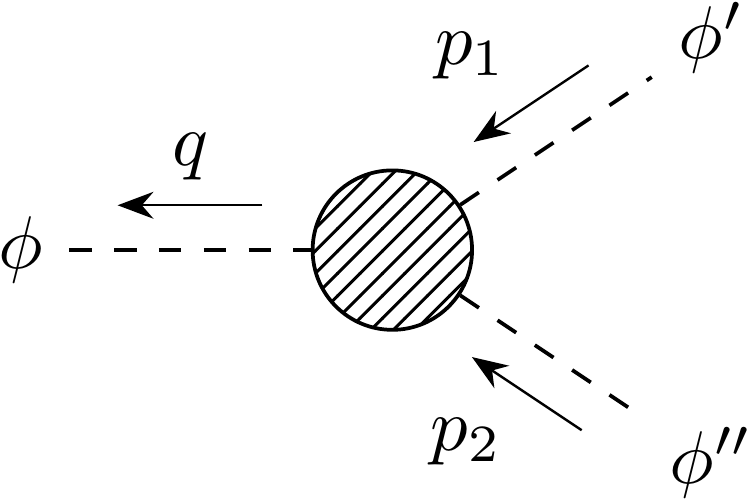}
    \caption{Momentum assignment for the renormalized $\phi ff^{\prime}$, $\phi VV^{\prime}$, $\phi V \phi^{\prime}$, and $\phi\phi^{\prime}\phi^{\prime\prime}$ vertices.}
    \label{fig: Feynman_diagrams}
\end{figure}

When we consider two body final states, $\phi$ can generally decay into a fermion and an anti-fermion $(\phi \to f\bar{f}')$, 
two gauge bosons ($\phi \to VV'$), a gauge and a scalar bosons $(\phi \to V\phi')$ and two scalar bosons ($\phi \to \phi' \phi''$). 
In order to discuss radiative corrections to the decay rate of these modes, it is convenient to introduce the following renormalized vertices 
$\hat{\Gamma}_{\phi ff'}$, $\hat{\Gamma}_{\phi VV'}^{\mu\nu}$, $\hat{\Gamma}_{\phi V\phi' }^\mu$ and $\hat{\Gamma}_{\phi \phi'\phi''}$. 
Except for $\hat{\Gamma}_{\phi \phi'\phi''}$, each renormalized vertex can be decomposed into the following form factors: 
\begin{align}
\hat{\Gamma}_{\phi ff'}(p_1^2,p_2^2,q^2)&=
\hat{\Gamma}_{\phi ff'}^{\rm S}+\gamma_5 \hat{\Gamma}_{\phi ff'}^{\rm P} + p_1\hspace{-3.5mm}/\hspace{2mm}\hat{\Gamma}_{\phi ff'}^{{\rm V}_1}
+p_2\hspace{-3.5mm}/\hspace{2mm}\hat{\Gamma}_{\phi ff'}^{{\rm V}_2}\notag\\
&\quad +p_1\hspace{-3.5mm}/\hspace{2mm}\gamma_5 \hat{\Gamma}_{\phi ff'}^{{\rm A}_1}
+p_2\hspace{-3.5mm}/\hspace{2mm}\gamma_5\hat{\Gamma}_{\phi ff'}^{{\rm A}_2}
+p_1\hspace{-3.5mm}/\hspace{2mm}p_2\hspace{-3.5mm}/\hspace{2mm}\hat{\Gamma}_{\phi ff'}^{\rm T}
+p_1\hspace{-3.5mm}/\hspace{2mm}p_2\hspace{-3.5mm}/\hspace{2mm}\gamma_5\hat{\Gamma}_{\phi ff'}^{\rm PT}, \label{eq:hff-form} \\
\hat{\Gamma}_{\phi VV'}^{\mu\nu}(p_1^2,p_2^2,q^2)&=g^{\mu\nu}\hat{\Gamma}_{\phi VV'}^1 + \frac{p_1^\nu p_2^\mu}{q^2}\hat{\Gamma}_{\phi VV'}^2 + i\epsilon^{\mu\nu\rho\sigma} \frac{p_{1\rho} p_{2\sigma}}{q^2} \hat{\Gamma}_{\phi VV'}^3,  \\
\hat{\Gamma}_{\phi V\phi' }^{\mu}(p_1^2,p_2^2,q^2)&=(p_1 + q)^\mu \hat{\Gamma}_{\phi V\phi' } ,
\end{align}
where $p_1^\mu$ and $p_2^\mu$ denote the incoming momenta, and $q^\mu$ represents the outgoing momentum of $\phi$, see Fig.~\ref{fig: Feynman_diagrams}.
Each renormalized form factor $\hat{\Gamma}_{\phi XY}$ can be further divided into the following terms: 
\begin{align}
\hat{\Gamma}^i_{\phi XY}(p_1^2,p_2^2,q^2)&=\Gamma^{i,{\rm tree}}_{\phi XY} + \Gamma^{i,{\rm loop}}_{\phi XY} ~~\text{with}~~ \Gamma^{i,{\rm loop}}_{\phi XY} = \Gamma^{i,{\rm 1PI}}_{\phi XY}+\delta\Gamma^{i}_{\phi XY},  \label{eq:form-loop}
\end{align}
where $\Gamma^{i,{\rm tree}}_{\phi XY}$, $\Gamma^{i,{\rm 1PI}}_{\phi XY}$ and $\delta\Gamma^{i}_{\phi XY}$ respectively denote contributions from tree diagrams, 1PI diagrams and counterterms. 

Let us also describe differences that can appear by using the different renormalization schemes in the HSM and the THDMs in order.
In the HSM, the scheme difference between the KOSY scheme and the PT scheme only appears in $\hat{\Gamma}_{\phi \phi' \phi''}$ as can be seen in Ref.~\cite{Kanemura:2017wtm}.  
The analytical expressions can be found in the reference. 
The scheme difference does not appear in the other vertices $\hat{\Gamma}_{\phi ff'}$, $\hat{\Gamma}_{\phi VV'}^{\mu\nu}$ and $\hat{\Gamma}_{\phi V\phi' }^\mu$. 
In the THDMs, similar to the HSM, the scheme difference between the KOSY1 scheme and the PT1 scheme only arises for the renormalized vertex functions $\Gamma_{\phi \phi' \phi''}$.
The scheme difference $\Delta \Gamma_{\phi \phi' \phi''}$ consists of the tadpole contributions, as given in Eqs.\eqref{eq:difhhh}, \eqref{eq:difbhhh} and \eqref{eq:difbhAA}. 
On the other hand, no scheme difference appears for $\Gamma_{\phi \phi' \phi''}$  between the KOSY2 scheme and the PT2 scheme, so that these two schemes give the same results for all the decay processes of $\phi$. 
In the end, three different results are obtained only for $\phi\to \phi' \phi''$ depending on these renormalization schemes in the THDMs.  
In Appendix.~\ref{app: schemediff}, we present the analytical expressions for the scheme difference. 

\subsection{Radiative corrections to the decay rate}

For decay processes that appear at the tree level, decay rates at NLO are generally expressed as 
\begin{align}
\Gamma(\phi \to XY)_{\rm NLO} &= \Gamma(\phi \to XY)_{\rm LO}[1 + \Delta_{\rm EW}(\phi \to XY) -\Delta r + \Delta_{\rm QCD}(\phi \to XY) ]  \notag \\
&+ \Gamma(\phi \to XY\gamma), 
\end{align}
where $\Gamma(\phi \to XY)_{\rm LO}$ is the decay rate at LO, $\Delta_{\rm EW}(\phi \to XY)$ is the electroweak correction,  $\Delta_{\rm QCD}(\phi \to XY)$ is the QCD correction and 
$\Gamma(\phi \to XY\gamma)$ is the contribution from the real photon emission which is required to guarantee IR divergence-free results. 
The contribution from the real photon emission vanishes when $\phi$, $X$, and $Y$ are electrically neutral. 
We separately show the contribution from the electroweak correction to the muon decay, $\Delta r$, which appears by the replacement of the VEV, i.e.,  $v^2 \to v^2(1 + \Delta r)$. 
The decay rates at LO are expressed as 
\begin{align}
\Gamma(\phi \to f\bar{f}')_{\rm LO} &= N_c^f\frac{m_\phi}{8\pi} \lambda^{1/2}(x_f,x_{f'})\notag\\
&\times \Big[(|\Gamma_{\phi ff'}^{\rm S,tree}|^2 + |\Gamma_{\phi ff'}^{\rm P,tree}|^2)(1-x_f - x_{f'}) -2(|\Gamma_{\phi ff'}^{\rm S,tree}|^2 - |\Gamma_{\phi ff'}^{\rm P,tree}|^2)\sqrt{x_fx_{f'}}\Big], \\
\Gamma(\phi \to VV')_{\rm LO} &= \frac{|\Gamma_{\phi VV'}^{1,{\rm tree}}|^2}{64\pi m_\phi(1 + \delta_{VV'})}\frac{\lambda(x_{V},x_{V'}) + 12x_Vx_{V'}}{x_Vx_{V'}}\lambda^{1/2}(x_{V},x_{V'}), \label{eq:phitoVVLO}\\
\Gamma(\phi \to V\phi')_{\rm LO} &= \frac{|\Gamma_{\phi V\phi'}^{\rm tree}|^2}{16\pi}\frac{m_{\phi}^3}{m_V^2}\lambda^{3/2}(x_{V},x_{\phi'}), \\
\Gamma(\phi \to \phi'\phi'')_{\rm LO} &= \frac{|\Gamma_{\phi \phi'\phi''}^{\rm tree}|^2}{16\pi m_\phi(1 + \delta_{\phi'\phi''})}\lambda^{1/2}(x_{\phi'},x_{\phi''}), 
\end{align}
% \begin{align}
% \Gamma(\phi \to \phi'\phi'')_{\rm LO} &= \frac{|\Gamma_{\phi \phi'\phi''}^{\rm tree}|^2}{16\pi m_\phi(1 + \delta_{\phi'\phi''})}\lambda^{1/2}(x_{\phi'},x_{\phi''}), \\
% \Gamma(\phi \to V\phi')_{\rm LO} &= \frac{|\Gamma_{\phi V\phi'}^{\rm tree}|^2}{16\pi}\frac{m_{\phi'}^3}{m_V^2}\lambda^{3/2}(x_{V},x_{\phi'}), \\
% \Gamma(\phi \to VV')_{\rm LO} &= \frac{|\Gamma_{\phi VV'}^{1,{\rm tree}}|^2}{64\pi m_\phi(1 + \delta_{VV'})}\frac{\lambda(x_{V},x_{V'}) + 12x_Vx_{V'}}{x_Vx_{V'}}\lambda^{1/2}(x_{V},x_{V'}), \\
% \Gamma(\phi \to f\bar{f}')_{\rm LO} &= N_c^f\frac{m_\phi}{8\pi} \lambda^{1/2}(x_f,x_{f'})\notag\\
% &\times \Big[(|\Gamma_{\phi ff'}^{\rm S,tree}|^2 + |\Gamma_{\phi ff'}^{\rm P,tree}|^2)(1-x_f - x_{f'}) -2(|\Gamma_{\phi ff'}^{\rm S,tree}|^2 - |\Gamma_{\phi ff'}^{\rm P,tree}|^2)\sqrt{x_fx_{f'}}\Big], 
% \end{align}
where $N_c^f=3~(1)$ for $f$ to be quarks (leptons), $x_a = m_a^2/m_\phi^2$ and $\lambda(x,y) = (1-x-y)^2-4xy$. In the above expressions, $V$ and $V'$ are either $W^\pm$ or $Z$~\footnote{For the CP-odd Higgs boson and the charged Higgs bosons, the decay $\phi \to VV'$ specified as $A\to ZZ/W^+W^-$ and $H^\pm \to W^\pm Z$, respectively. Since they are loop-induced processes, the analytical expressions of the decay rates are presented by Eq.~\eqref{eq:phiVV_loopind}.}.
The electroweak corrections $\Delta_{\rm EW}$ are expressed for each process as
\begin{align}
\Delta_{\rm EW}(\phi \to f\bar{f}') & = 
\Big[(|\Gamma_{\phi ff'}^{\rm S,tree}|^2 + |\Gamma_{\phi ff'}^{\rm P,tree}|^2)(1-x_f - x_{f'}) -2(|\Gamma_{\phi ff'}^{\rm S,tree}|^2 - |\Gamma_{\phi ff'}^{\rm P,tree}|^2)\sqrt{x_fx_{f'}}\Big]^{-1} \notag \\
&\times \Big[2\left\{\text{Re}[\Gamma_{\phi ff'}^{\rm S,tree}\Gamma_{\phi ff'}^{{\rm loop}*}]+\text{Re}[\Gamma_{\phi ff'}^{\rm P,tree}\tilde{\Gamma}_{\phi ff'}^{{\rm loop}*}] \right\}(1 - x_f - x_{f'})  \notag\\
& - 4\left\{\text{Re}[\Gamma_{\phi ff'}^{\rm S,tree}\Gamma_{\phi ff'}^{{\rm loop}*}] - \text{Re}[\Gamma_{\phi ff'}^{\rm P,tree}\tilde{\Gamma}_{\phi ff'}^{{\rm loop}*}] \right\}\sqrt{x_fx_{f'}} \Big]+ \Delta_{\rm mix}(\phi \to f\bar{f}'), \\
\Delta_{\rm EW}(\phi \to VV') & = 
\frac{2\text{Re}[\Gamma_{\phi VV'}^{1,{\rm tree}}\Gamma^{{1,{\rm loop}}*}_{\phi VV'}]}{|\Gamma^{1,{\rm tree}}_{\phi VV'}|^2} + \frac{\text{Re}[\Gamma_{\phi VV'}^{1,{\rm tree}}\Gamma^{{2,{\rm loop}}*}_{\phi VV'}]}{|\Gamma^{1,{\rm tree}}_{\phi VV'}|^2}
\frac{(1-x_V - x_{V'})\lambda(x_V,x_{V'}) }{\lambda(x_V,x_{V'}) + 12x_Vx_{V'}}\notag\\
&  - \text{Re}\hat{\Pi}_{VV}'(m_V^2) -  \text{Re}\hat{\Pi}_{V'V'}'(m_{V'}^2) , \\
\Delta_{\rm EW}(\phi \to V\phi') &= \frac{2\text{Re}[\Gamma^{\rm tree}_{\phi V\phi'}\Gamma^{{\rm loop}*}_{\phi V\phi'}]}{|\Gamma^{\rm tree}_{\phi V\phi'}|^2} - \text{Re}\hat{\Pi}_{VV}'(m_V^2) + \Delta_{\rm mix}(\phi \to V\phi'), \\ 
\Delta_{\rm EW}(\phi \to \phi'\phi'') &= \frac{2\text{Re}[\Gamma^{\rm tree}_{\phi\phi'\phi''}\Gamma^{{\rm loop}*}_{\phi\phi'\phi''}]}{|\Gamma^{\rm tree}_{\phi\phi'\phi''}|^2} + \Delta_{\rm mix}(\phi \to \phi'\phi'') , 
\end{align}
% \begin{align}
% \Delta_{\rm EW}(\phi \to \phi'\phi'') &= \frac{2\text{Re}[\Gamma^{\rm tree}_{\phi\phi'\phi''}\Gamma^{{\rm loop}*}_{\phi\phi'\phi''}]}{|\Gamma^{\rm tree}_{\phi\phi'\phi''}|^2}  , \\
% \Delta_{\rm EW}(\phi \to V\phi') &= \frac{2\text{Re}[\Gamma^{\rm tree}_{\phi V\phi'}\Gamma^{{\rm loop}*}_{\phi V\phi'}]}{|\Gamma^{\rm tree}_{\phi V\phi'}|^2} - \text{Re}\hat{\Pi}_{VV}'(m_V^2) + \Delta_{\rm mix}(\phi \to V\phi'), \\ 
% %
% \Delta_{\rm EW}(\phi \to VV') & = 
% \frac{2\text{Re}[\Gamma_{\phi VV'}^{1,{\rm tree}}\Gamma^{{1,{\rm loop}}*}_{\phi VV'}]}{|\Gamma^{1,{\rm tree}}_{\phi VV'}|^2} + \frac{\text{Re}[\Gamma_{\phi VV'}^{1,{\rm tree}}\Gamma^{{2,{\rm loop}}*}_{\phi VV'}]}{|\Gamma^{1,{\rm tree}}_{\phi VV'}|^2}
% \frac{(1-x_V - x_{V'})\lambda(x_V,x_{V'}) }{\lambda(x_V,x_{V'}) + 12x_Vx_{V'}}\notag\\
% &  - \text{Re}\hat{\Pi}_{VV}'(m_V^2) -  \text{Re}\hat{\Pi}_{V'V'}'(m_{V'}^2) , \\ 
% %
% \Delta_{\rm EW}(\phi \to f\bar{f}') & = \Big[2\left\{\text{Re}[\Gamma_{\phi ff'}^{\rm S,tree}\Gamma_{\phi ff'}^{{\rm loop}*}]+\text{Re}[\Gamma_{\phi ff'}^{\rm P,tree}\tilde{\Gamma}_{\phi ff'}^{{\rm loop}*}] \right\}(1 - x_f - x_{f'})  \notag\\
% & - 4\left\{\text{Re}[\Gamma_{\phi ff'}^{\rm S,tree}\Gamma_{\phi ff'}^{{\rm loop}*}] - \text{Re}[\Gamma_{\phi ff'}^{\rm P,tree}\tilde{\Gamma}_{\phi ff'}^{{\rm loop}*}] \right\}\sqrt{x_fx_{f'}} \Big]\notag\\
% &\times \Big[(|\Gamma_{\phi ff'}^{\rm S,tree}|^2 + |\Gamma_{\phi ff'}^{\rm P,tree}|^2)(1-x_f - x_{f'}) -2(|\Gamma_{\phi ff'}^{\rm S,tree}|^2 - |\Gamma_{\phi ff'}^{\rm P,tree}|^2)\sqrt{x_fx_{f'}}\Big]^{-1}, 
% \end{align}
where
\begin{align}
\Gamma_{\phi ff'}^{\rm loop} &= \Gamma_{\phi ff'}^{\rm S,loop} + m_{f'}\Gamma_{\phi ff'}^{{\rm V}_1,{\rm loop}} - m_{f}\Gamma_{\phi ff'}^{{\rm V}_2,{\rm loop}}
+(m_\phi^2+m_fm_{f'} -m_f^2 - m_{f'}^2)\Gamma_{\phi ff'}^{\rm T,loop}, \\
\tilde{\Gamma}_{\phi ff'}^{\rm loop} &= \Gamma_{\phi ff'}^{\rm P,loop} - m_{f'}\Gamma_{\phi ff'}^{{\rm A}_1,{\rm loop}} - m_{f}\Gamma_{\phi ff'}^{{\rm A}_2,{\rm loop}}
+(m_\phi^2-m_fm_{f'} -m_f^2 - m_{f'}^2)\Gamma_{\phi ff'}^{\rm PT,loop}. 
\end{align}
The external-leg corrections to the on-shell gauge boson ${\rm Re}\hat{\Pi}^\prime_{VV} (m_V^2)$ appear in $\Delta_{\rm EW}(\phi \to V\phi')$ and $\Delta_{\rm EW}(\phi \to VV')$, because 
the residue of the self-energy for the on-shell weak gauge boson is not set to unity in the applied renormalization scheme for electroweak parameters (see, e.g., Ref.~\cite{Hollik:1988ii}). 
%The correction factor $\Delta_{\rm mix}(\phi \to XY)$ corresponds to the non-vanishing contributions from mixing self-energies between two boson states.  
%This type of contribution arises if $\phi$, $\phi'$ and/or $\phi''$ is the charged Higgs bosons in the THDM, where $\hat{\Pi}_{H^+G^-}(m_{H^\pm}^2)=0$ is not imposed as the renormalization condition in our renormalization scheme~\cite{Kanemura:2015mxa}. 
%The analytical expression for $\Delta_{\rm mix}(\phi \to V\phi')$ is given in Ref.~\cite{Aiko:2021can} and Appendix~\ref{app: phiphiphi}. 
The correction factor $\Delta_{\rm mix}(\phi \to XY)$ corresponds to the non-vanishing contributions from mixing self-energies between two boson states.  
This type of contribution arises if external lines contain the charged Higgs boson in the THDM, where $\hat{\Pi}_{H^+G^-}(m_{H^\pm}^2)=0$ is not imposed as the renormalization condition in our renormalization scheme~\cite{Kanemura:2015mxa}. 
The analytical expressions are given in Ref.~\cite{Aiko:2021can} for $\Delta_{\rm mix}(\phi \to f\bar{f}')$ and $\Delta_{\rm mix}(\phi \to V\phi')$, and Appendix~\ref{app: phiphiphi} for $\Delta_{\rm mix}(\phi \to \phi'\phi'')$. 

In Table~\ref{tab:zeta}, we summarize the references giving the analytical expressions for 1PI vertices, counterterms, and real photon emissions, which are needed to compute the two-body decay rates
of all the additional Higgs bosons.   

\begin{table}[t]
\begin{center}
{\renewcommand\arraystretch{1.2}
\begin{tabular}{l|cccc}\hline\hline
Modes                    & 1PI vertex & Counterterm & Real emission & Models \\\hline
$H \to f\bar{f}$    &  \cite{Kanemura:2015mxa} & \cite{Kanemura:2022ldq} & \cite{Kanemura:2022ldq} & HSM\\\hline
$H \to W^+W^-$      &  \cite{Kanemura:2015mxa} & \cite{Kanemura:2022ldq} & \cite{Aiko:2022gmz} & HSM\\\hline
$H \to ZZ$          &  \cite{Kanemura:2015mxa} & \cite{Kanemura:2022ldq} & - & HSM \\\hline
$H \to AZ$          &  \cite{Aiko:2022gmz} & \cite{Aiko:2022gmz} & - & IDM\\\hline
$H \to W^\pm H^\mp$  &  \cite{Aiko:2021can} & \cite{Aiko:2021can} & \cite{Aiko:2021can} & HSM\\\hline
$H \to hh$    &\cite{Kanemura:2022ldq} & \cite{Kanemura:2022ldq} & - & HSM\\\hline
$H \to AA$    & App.~\ref{app: phiphiphi} & App.~\ref{app: phiphiphi}  & - & IDM ($h\to AA $ (App.~\ref{app: phiphiphi}))\\\hline
$H \to H^+H^-$    & App.~\ref{app: phiphiphi} &  App.~\ref{app: phiphiphi} & App.~\ref{app:real emission} & IDM ($h\to H^+H^-$ (App.~\ref{app: phiphiphi}))\\\hline
$H \to \gamma\gamma/Z\gamma/gg$    & \cite{Kanemura:2015mxa} &-  &-  & HSM\\\hline\hline
$A \to f\bar{f}$    &\cite{Aiko:2022gmz}&\cite{Aiko:2022gmz} & \cite{Aiko:2022gmz} & -\\\hline
$A \to Zh$    &\cite{Aiko:2022gmz}&\cite{Aiko:2022gmz} &- & - \\\hline
$A \to ZH$    &\cite{Aiko:2022gmz}&\cite{Aiko:2022gmz} &-& IDM \\\hline
$A \to W^\pm H^\mp   $ &\cite{Aiko:2021can} & \cite{Aiko:2022gmz}  & \cite{Aiko:2022gmz} & IDM\\\hline
$A \to \gamma\gamma/Z\gamma/gg/WW/ZZ   $ &\cite{Aiko:2022gmz}&-&-&- \\\hline\hline
$H^\pm \to f\bar{f}'$    &\cite{Aiko:2021can}&\cite{Aiko:2021can} & \cite{Aiko:2021can} & -\\\hline
$H^\pm \to W^\pm h$    &\cite{Aiko:2021can}&\cite{Aiko:2021can} & \cite{Aiko:2021can} &  -\\\hline
$H^\pm \to W^\pm H/W^\pm A$    &\cite{Aiko:2021can}&\cite{Aiko:2021can} & \cite{Aiko:2021can} & IDM \\\hline
$H^\pm \to W^\pm Z$    & \cite{Aiko:2021can} & \cite{Aiko:2021can} & - &- \\\hline
$H^\pm \to W^\pm \gamma $    &\cite{Aiko:2021can} &-  & -& -  \\\hline
\end{tabular}}
\cprotect\caption{References for the expressions of each contribution to the decay rate in the THDMs. 
In Ref.~\cite{Kanemura:2015mxa}, the expression for the $h$ vertices is presented, so that the corresponding expression for the $H$ vertices is obtained by taking the appropriate replacement of the coupling for $h$
with the corresponding one for $H$.
The column ``Models'' shows the possible other models to give the corresponding decay mode, which are implemented in \verb|H-COUP_3.0|, i.e., the IDM and the HSM.}
\label{tab:zeta}
\end{center}
\end{table}

%We note that the decay rate for the $\phi \to f\bar{f}'$ process can be simplified for a CP-even Higgs boson ($\phi = \phi_{\rm even}$) and a CP-odd ($\phi = \phi_{\rm odd}$) with $f' = f$ as follows: 
%\begin{align}
%\Gamma(\phi \to f\bar{f}')_{\rm LO} &= \frac{m_\phi}{8\pi} \lambda^{1/2}(x_f,x_{f'})\notag\\
%&\times \Big[(|\Gamma_{\phi ff'}^{\rm S,tree}|^2 + |\Gamma_{\phi ff'}^{\rm P,tree}|^2)(1-x_f - x_{f'}) -2(|\Gamma_{\phi ff'}^{\rm S,tree}|^2 - |\Gamma_{\phi ff'}^{\rm P,tree}|^2)\sqrt{x_fx_{f'}}\Big], 
%\end{align}

For decay processes induced at loop levels, their decay rates are given at LO as 
\begin{align}
\Gamma(\phi \to V\gamma)_{\rm LO} &= \frac{(1 - x_V)^3}{32\pi m_\phi(1 + \delta_{V\gamma}) }(|\hat{\Gamma}_{\phi V\gamma}^{2,{\rm loop}}|^2 + |\hat{\Gamma}_{\phi V\gamma}^{3,{\rm loop}}|^2 ), \quad (V=\gamma,~Z,~W^\pm)\\
\Gamma(\phi \to gg)_{\rm LO}      &= \frac{1}{8\pi m_\phi }(|\hat{\Gamma}_{\phi gg}^{2,{\rm loop}}|^2 + |\hat{\Gamma}_{\phi gg}^{3,{\rm loop}}|^2 ). 
\end{align}
In the above expression, we applied the Ward identity by which the contribution from $\hat{\Gamma}_{\phi XY}^{1,{\rm loop}}$ is rewritten by $\hat{\Gamma}_{\phi XY}^{2,{\rm loop}}$. 
If $\phi \to VV'$ decays appear at the one-loop level, e.g., $H^\pm \to W^\pm Z$ in the THDMs, the decay rate is expressed at LO as~
%\footnote{If one specifies $\phi \to VV'$ as  $H \to ZZ/W^+W^-$ in the HSM or the THDM, one can obtain  $\Gamma(H \to ZZ/W^+W^-)_{\rm LO}$ in the alignment limit, i.e., $\Gamma_{\phi VV'}^{1,{\rm tree}}=0$, where $H \to ZZ/W^+W^-$ are loop-induced. }
\footnote{If one specifies $\phi \to VV'$ as  $H \to ZZ/W^+W^-$ in the HSM or the THDM, these processes realize at the tree level. The analytical expressions are given by Eq.~\eqref{eq:phitoVVLO}. }
~\footnote{We note that, for the charged Higgs decays $H^\pm\to W^\pm Z$ in the THDM, $H^\pm G^\mp$ and  $H^\pm W^\mp$ mixing contributions exist. They are included in $\hat{\Gamma}^{1,{\rm loop}}_{\phi VV'}$.}
\begin{align} \label{eq:phiVV_loopind}
\Gamma(\phi \to VV')_{\rm LO} & = \frac{\lambda^{3/2}(x_V,x_{V'})}{64\pi m_\phi x_Vx_{V'}}\Bigg\{|\hat{\Gamma}_{\phi VV'}^{1,{\rm loop}}|^2\left[1 + \frac{12 x_Vx_{V'}}{\lambda(x_V,x_{V'})}\right] 
+ \frac{|\hat{\Gamma}_{\phi VV'}^{2,{\rm loop}}|^2 }{4}\lambda(x_V,x_{V'}) \notag\\
& + 2x_Vx_{V'}|\hat{\Gamma}_{\phi VV'}^{3,{\rm loop}}|^2 + \text{Re}[\hat{\Gamma}_{\phi VV'}^{1,{\rm loop}}\hat{\Gamma}_{\phi VV'}^{2,{\rm loop}*}](1 - x_V - x_{V'}) \Bigg\}. 
\end{align}
% If $\phi \to W^\pm Z$ decays appear at the one-loop level, e.g., the THDMs, the decay rate is expressed at LO as
% \begin{align}
% \Gamma(\phi \to W^\pm Z)_{\rm LO} & = \frac{\lambda^{3/2}(x_W,x_Z)}{64\pi m_\phi x_Wx_Z}\Bigg\{|\hat{\Gamma}_{\phi WZ}^{1,{\rm loop}}|^2\left[1 + \frac{12 x_Wx_Z}{\lambda(x_W,x_Z)}\right] 
% + \frac{|\hat{\Gamma}_{\phi WZ}^{2,{\rm loop}}|^2 }{4}\lambda(x_W,x_Z) \notag\\
% & + 2x_Wx_Z|\hat{\Gamma}_{\phi WZ}^{3,{\rm loop}}|^2 + \text{Re}[\hat{\Gamma}_{\phi WZ}^{1,{\rm loop}}\hat{\Gamma}_{\phi WZ}^{2,{\rm loop}*}](1 - x_W - x_Z) \Bigg\}. 
% \end{align}

The QCD corrections are included in the following processes,
\begin{align} \label{eq:QCD2body}
    \phi\to q\bar{q}^\prime,\quad
    \phi\to t\bar{t},\quad
    \phi\to t\bar{b},\quad
    \phi\to \gamma\gamma,\quad
    \phi\to Z\gamma,\quad
    \phi\to gg.
\end{align}
 {The evaluation of QCD corrections for the processes of Eq.~\eqref{eq:QCD2body} is common for all the models.} 
% \textcolor{blue}{if the LO process exists. 
% We note that in the extended Higgs models presented in Sec.~\ref{sec: model}, the additional Higgs bosons do not have the color charge, so that the evaluation of QCD corrections for the processes of Eq.~\eqref{eq:QCD2body} is common for all the models. } 
In the following paragraphs, we briefly describe
how the QCD corrections are included for each decay mode in \verb|H-COUP_3.0|.
See Ref.~\cite{Aiko:2020ksl} for detailed expressions for these QCD corrections.

For the decays into light quarks $\phi\to q\bar{q}^\prime$, 
the corrections up to next-to-next-to leading order (NNLO) are computed in the $\overline{\rm MS}$ scheme~\cite{Mihaila:2015lwa, Gorishnii:1990zu,Gorishnii:1991zr,Chetyrkin:1995pd,Larin:1995sq}. 
The NNLO corrections involve top-quark loop contributions evaluated in the heavy top-quark mass limit $m_{\phi}\ll m_t $. 
For the decays including the top-quark $\phi\to t\bar{t}$ and $\phi\to t\bar{b}$, the dominant QCD corrections would depend on the size of the mass of additional Higgs bosons $m_{\phi}$.  
If these masses are taken to be around the threshold region (e.g., $m_{H,A}^{} \sim 2m_t$ and $m_{H^\pm} \sim m_t+m_b$) the corrections with top-quark mass are important, which is evaluated at NLO in the on-shell scheme~\cite{Drees:1989du, Djouadi:2005gj, Djouadi:1994gf}. 
Conversely, if these are far above the threshold region (e.g., $m_{H,A}^{} \gg 2m_t$ and $m_{H^\pm}^{} \gg m_t+m_b$), the logarithmic contributions $\log (m_{\phi}^2/m_t^2)$ become important. 
One can evaluate the corrections including such logarithmic contributions up to NNLO in the $\overline{\rm MS}$ scheme in the same way as $\phi\to q\bar{q}$. 
Since in the intermediate region of $m_{\phi}$, both above contributions could have a large influence, we interpolate them according to Ref.~\cite{Djouadi:1997yw}. 

We evaluate QCD corrections to  $\phi\to \gamma\gamma$~\cite{Dawson:1993qf,Spira:1995rr,Harlander:2005rq} and $\phi\to Z\gamma$~\cite{Djouadi:2005gj} at NLO. 
While the analytical formulae of the former can be applied to any mass region of additional Higgs bosons, those in the heavy top-quark limit are applied to the latter. 
To moderate these corrections, we choose the renormalization scale $\mu=m_{\phi}/2$ only for $\phi\to \gamma\gamma$ and $\phi \to Z\gamma$ ($\mu=m_{\phi}$ is used for the other processes.).
For $\phi\to gg$, the corrections are calculated up to NNLO. 
The NLO corrections are composed of the virtual gluon loop corrections and the real emissions $\phi\to ggg $ and $\phi\to gq\bar{q}$. 
Following Ref.~\cite{Harlander:2005rq}, we implement these contributions evaluated in the heavy top-quark limit. 
We neglect the remaining contributions for the real emissions, which are generally smaller than the virtual corrections (see the detail in Ref.~\cite{Spira:1995rr}). 
For the NNLO QCD corrections to $\phi\to gg$, we use the formula evaluated in the heavy top-quark limit~\cite{Chetyrkin:1997iv, Chetyrkin:1998mw}. 

As described above, some of the QCD corrections are computed by taking the heavy top-quark limit. 
However, additional Higgs bosons can be heavier than the top-quark, so that the QCD corrections derived in the heavy top-quark limit might be unreliable in the case of $m_\phi \gtrsim m_t$. 
For instance, the top-quark loop contributions to $\phi\to q\bar{q}$ contain the logarithmic contributions $\log (m_{\phi}^2/m_t^2)$, which gives sizable contributions in the case of $m_\phi \gg m_t$. 
%Hence, we only include the contributions of the heavy top-quark limit containing such logarithmic contributions, i.e., the top loop contributions to $\phi\to q\bar{q}$, the NNLO corrections to $\phi\to gg$ in the evaluation of the QCD corrections if a heavy Higgs boson is lighter than the top quark ($m_{\phi}<m_t$).
Hence, we only include the contributions given in the heavy top-quark limit, i.e., the top-loop contributions to $\phi\to q\bar{q}$, the real emission contributions at NLO to $\phi\to gg$ and the NNLO corrections to $\phi\to gg$ in the evaluation of the QCD corrections if an additional Higgs boson is lighter than the top-quark ($m_{\phi}<m_t$).

Apart from the QCD correction factor $\Delta_{\rm QCD}(\phi\to XY)$, quark masses coming from the Yukawa couplings are replaced by the running quark masses in $\Gamma_{\rm LO}(\phi\to XY )$ for all the processes in Eq.~\eqref{eq:QCD2body}. 

The decays into an off-shell gauge boson $\phi \to VV^\ast \to Vf\bar{f}$ and $\phi \to \phi^\prime V^\ast \to \phi^\prime f\bar{f}$ can happen in the case of  $m_\phi < 2m_V$ for the former and $m_\phi < m_{\phi^\prime} + m_V^{}$ for the latter. 
While they are evaluated at electroweak LO~\footnote{For the off-shell decays of the 125 GeV Higgs boson $h$, $h \to VV^\ast \to V f\bar{f}$, the NLO electroweak corrections are included. }, we incorporate NLO QCD corrections to them~\cite{Albert:1979ix} in \verb|H-COUP_3.0|.

\section{Program description} \label{sec: structure}

One can download \verb|H-COUP_3.0| and the previous versions from the following web page,\\
\vspace{-0.6cm}
\begin{center}
\url{http://www-het.phys.sci.osaka-u.ac.jp/~hcoup}.
\end{center}
In order to run the \texttt{H-COUP} program, a \texttt{Fortran} compiler (\texttt{gfortran} is recommended) and {\tt LoopTools}~\cite{Hahn:1998yk} are required.
The installation and running procedures are the same as \verb|H-COUP_2.0|, and these are described in the previous manual~\cite{Kanemura:2019slf}, except for the procedure to specify the path of \verb|LoopTools|.
Open Makefile by an editor and replace \verb|PATH_TO_libooptools_a| and \verb|PATH_TO_looptools_h| appearing in the lines \verb|LPATHa| and \verb|LPATHh| with the correct paths to the \verb|libooptools.a| and \verb|looptools.h|, respectively.
In the following, we briefly overview the structure of \verb|H-COUP_3.0| and explain the improvement from previous versions.

The \texttt{H-COUP} program is composed of three blocks; input, computation, and output blocks~\cite{Kanemura:2017gbi, Kanemura:2019slf}.
In the input block, \verb|H-COUP_3.0| reads the input files in \verb|$HCOUP-3.0/inputs|, where \verb|$HCOUP-3.0| indicates a path of the \verb|HCOUP-3.0| directory.
In \verb|H-COUP_3.0|, the model and the order of calculations are specified in \verb|in_main.txt|.
This part is different from \verb|H-COUP_2.0|, where one specifies them from the command line interface.
The model IDs are defined as \verb|HSM=1|, \verb|THDM-I=2|, \verb|THDM-II=3|, \verb|THDM-X=4|, \verb|THDM-Y=5|, and \verb|IDM=6|.
The order of electroweak corrections is specified by \verb|LO=0| or \verb|NLO=1|, while the order of QCD corrections are defined as \verb|LO(quark mass:OS)=-1|, \verb|LO(quark mass:MSbar)=0|, \verb|NLO=1|, and \verb|NLO=2|.
The option, \verb|LO(quark mass:OS)|, provides the LO results calculated with OS quark masses, while \verb|LO(quark mass:MSbar)| performs LO calculation with $\overline{\rm MS}$ quark masses.
The detailed descriptions of the SM-like Higgs boson decays are given in Ref.~\cite{Kanemura:2019kjg}.
For those of the heavy Higgs boson decays, see Refs.~\cite{Kanemura:2022ldq, Aiko:2021can, Aiko:2022gmz}.
The example of the \verb|in_main.txt| is shown in List.~\ref{file_in_main}.

\begin{table}[t]
\begin{tabular}{|llll|}\hline
Parameter                     & Definition in {\tt H-COUP} & Description                   & Default value   \\\hline \hline 
  $\alpha_{\text{em}}$        & \verb|alpha_em|            & Fine structure constant       & $137.035999084^{-1}$ \\\hline 
  $m_Z$                       & \verb|mz|                  & $Z$ mass                      & 91.1876 GeV \\\hline
  $G_F$                       & \verb|G_F|                 & Fermi constant                & 1.1663788$\times 10^{-5}$ GeV$^{-2}$ \\\hline
  $\Delta \alpha_{\text{em}}$ & \verb|del_alpha|           & Shift of $\alpha_{\text{em}}$ & 0.06627     \\\hline 
  $\alpha_s(m_Z^{})$          & \verb|alpha_s|             & Strong coupling               & 0.1179      \\\hline 
  $m_h$                       & \verb|mh|                  & Higgs boson mass              & 125.25 GeV   \\\hline
  $m_t$                       & \verb|mt|                  & On-shell $t$ mass             & 172.5 GeV   \\\hline
  $m_b$                       & \verb|mb|                  & On-shell $b$ mass             & 4.78 GeV    \\\hline 
  $m_c$                       & \verb|mc|                  & On-shell $c$ mass             & 1.67 GeV    \\\hline 
  $\overline{m}_b(m_b)$       & \verb|mb_ms|               & $\overline{\rm MS}$ $b$ mass  & 4.18 GeV    \\\hline 
  $\overline{m}_c(m_c)$       & \verb|mc_ms|               & $\overline{\rm MS}$ $c$ mass  & 1.27 GeV    \\\hline 
  $m_\tau$                    & \verb|mtau|                & $\tau$ mass                   & 1.77686 GeV \\\hline
  $m_\mu$                     & \verb|mmu|                 & $\mu$ mass                    & 0.1056583755 GeV  \\\hline
 \end{tabular}
\caption{Input global SM parameters. All these parameters are defined by double precision. All input values are taken from particle data group~\cite{ParticleDataGroup:2022pth}.}
\label{global1}
\end{table}

The model-independent parameters are read from the input files, \verb|in_sm.txt| and \verb|momentum.txt|.
The SM parameters listed in Table~\ref{global1} are specified in \verb|in_sm.txt|.
The squared momenta of the renormalized form factors of the SM-like Higgs bosons are read from \verb|momentum.txt|.
The detailed descriptions of these parameters can be found in Refs.~\cite{Kanemura:2017gbi, Kanemura:2019slf}.
If one changes the inputs in \verb|in_sm.txt|, execute
\begin{center}
\fbox{\$  make clean}
\end{center}
before numerical evaluations.

The model-dependent parameters in the HSM, THDM, and IDM are specified in \verb|in_hsm.txt|, \verb|in_thdm.txt|, and \verb|in_idm.txt|, respectively.
The input parameters and their default values are summarized in Tables~\ref{input_hsm}, \ref{input_thdm}, and \ref{input_idm}.
The cutoff scale $\Lambda$ is used for the theoretical constraints, such as triviality and vacuum stability bounds.
In addition, we have newly introduced the renormalization scale \verb|mu_r|, which is used to fix the $\overline{\rm MS}$ renormalized quantities.
In the HSM and THDM, we have the \verb|SCHEME| option, which specifies the renormalization scheme of the tadpole and dimensionful parameters, which was discussed in Sec.~\ref{sec: reno}.
The example of the \verb|in_hsm.txt| is shown in List.~\ref{file_in_hsm}.

\begin{table}[t]
\begin{tabular}{|c||c c c c c c c c|}\hline
                  & \multicolumn{8}{c|}{HSM}     \\\hline\hline
Parameters        & $m_H^{}$ & $\alpha$  & $\mu_S^{}$ & $\lambda_S$ & $\lambda_{\Phi S}^{}$ & $\Lambda$ & $\mu_{R}$ & Scheme \\\hline 
{\tt H-COUP} def. & \verb|mbh| & \verb|alpha| & \verb|mu_s| & \verb|lam_s| & \verb|lam_phis| & \verb|cutoff| & \verb|mu_r| & \verb|SCHEME| \\\hline
Default value     & 500 GeV     & 0.1    & 0    & 0.1     & 0     & 3 TeV & 500 GeV & 0 \\\hline
\end{tabular}
\caption{Input parameters in the HSM.  All parameters are defined by double precision.}
\label{input_hsm}
\vspace{5mm}
\begin{tabular}{|c||c c c c c c c c c c|}\hline
                  & \multicolumn{10}{c|}{THDM}     \\\hline\hline
Parameters        &  $m_{H^\pm}^{}$ & $m_A^{}$  & $m_H^{}$   & $M^2$         & $s_{\beta-\alpha}$ & $\text{Sign}(c_{\beta-\alpha})$  & $\tan\beta$ & $\Lambda$     & $\mu_{R}$ & Scheme \\\hline 
{\tt H-COUP} def. & \verb|mch|      & \verb|ma| & \verb|mbh| & \verb|bmsq|   & \verb|sin_ba|      & \verb|sign|                         & \verb|tanb| & \verb|cutoff| & \verb|mu_r| & \verb|SCHEME| \\\hline 
Default value     &  500 GeV        & 500 GeV   & 500 GeV    & (450 GeV)$^2$ & 1                  & 1                                   & 1.5         & 3 TeV         & 500 GeV     &  0\\\hline
\end{tabular}
\caption{Input parameters in the THDMs. 
All parameters are defined by double precision except for {\sf sign} which is defined by an integer and can be either 1 or $-1$. }
\label{input_thdm}
\vspace{5mm}
\begin{tabular}{|c||c c c c c c c|}\hline
                  & \multicolumn{7}{c|}{IDM}     \\\hline\hline
Parameters        &  $m_{H^\pm}^{}$ & $m_A^{}$  & $m_H^{}$   & $\mu_2^2$      & $\lambda_2$ & $\Lambda$ & $\mu_{R}$\\\hline 
{\tt H-COUP} def. &  \verb|mch|     & \verb|ma| & \verb|mbh| &  \verb|mu2sq|  & \verb|lam2| & \verb|cutoff| & \verb|mu_r| \\\hline 
Default value     &  500 GeV        & 500 GeV   & 500 GeV    &  (500 GeV)$^2$ & 0.1         & 3 TeV         & 500 GeV\\\hline
\end{tabular}
\caption{Input parameters in the IDM. All parameters are defined by double precision.}
\label{input_idm}
\end{table}

In the computation block, \verb|H-COUP_3.0| calculates the decay rates of any Higgs bosons in the specified model at a given order of calculations, where the evaluation of the loop functions are performed with \texttt{LoopTools}~\cite{Hahn:1998yk}.
The possible decay modes of the additional Higgs bosons are listed in Table~\ref{tab:zeta}\footnote{In several modes such as $H\to hh$ in the THDM, partial decay widths would take negative values depending on the input parameters due to the truncation of two-loop effects (see Refs.~\cite{Kanemura:2022ldq, Aiko:2021can, Aiko:2022gmz}).
In the current version, we simply output negative values showing a warning message in a command line interface.}.

In the output block, \verb|H-COUP_3.0| creates output files in \verb|$HCOUP-3.0/outputs|.
In addition to the \verb|out_sm.txt|, \verb|outGamma_sm.txt| and \verb|outBR_sm.txt|, one has the output files for specifying models, e.g., \verb|out_hsm.txt|, \verb|outGamma_hsm.txt| and \verb|outBR_hsm.txt| for the HSM.
In the following, we represent them as \verb|out_xx.txt|, \verb|outGamma_xx.txt|, and \verb|outBR_xx.txt| for simplicity.
In \verb|out_xx.txt|, the values of the renormalized form factors are listed.
The predictions for the partial decay rates of Higgs bosons and their total widths are given in \verb|outGamma_xx.txt| with the values of the input parameters.
Similarly, the predictions for the decay branching ratios are given in \verb|outBR_xx.txt|.
One can check whether a given parameter set is allowed or excluded under the constraints, such as perturbative unitarity, vacuum stability, triviality, true vacuum conditions, and/or electroweak precision tests.
The examples of \verb|outGamma_hsm.txt| and \verb|outBR_hsm.txt| are shown in Lists.~\ref{file_outGamma} and \ref{file_outBR}, which are calculated by using the inputs given in Lists.~\ref{file_in_main} and \ref{file_in_hsm}.

% \vskip\baselineskip

\begin{lstlisting}[caption=Example of the input file (in\_main.txt),label=file_in_main,frame=lines]
!===============================!
!                               !
! Input parameters for main     !
!                            	!
!===============================!

  1     ! Model ID: 1 = HSM, 2 = THDM-I, 3 = THDM-II, 4 = THDM-X, 5 = THDM-Y, 6 = IDM
  1     ! Order of EW: 0 = LO, 1 = NLO
  2     ! Order of QCD: -1 = LO(quark mass:OS), 0 = LO(quark mass:MSbar), 1 = NLO, 2 = NNLO
\end{lstlisting}

\begin{lstlisting}[caption=Example of the input file (in\_hsm.txt),label=file_in_hsm,frame=lines]
!================================!
!                                !
! Input parameters for the HSM   !
!                                !
!================================!

  500.d0           ! m_H in GeV
  0.1d0            ! alpha
  0.d0             ! lambda_{phi S}
  0.1d0            ! lambda_S
  0.d0             ! mu_S in GeV
  3.d3             ! cutoff in GeV
  400.d0           ! MSbar renormalization scale in GeV
  0                ! 0: Pinched tadpole, 1: KOSY with PT  
\end{lstlisting}

\begin{lstlisting}[caption=Example of the output file (outGamma\_hsm.txt),label=file_outGamma,frame=lines]
#======================================================================
#
#               H-COUP [Version 3.0 (Novemver 28, 2023)]
#      Program for full-NLO predictions of any Higgs-boson decays
#                     in non-minimal Higgs models
#
#             http://www-het.phys.sci.osaka-u.ac.jp/~hcoup
#
#======================================================================
BLOCK MODEL #
     1      1   # HSM
     2      0   # MSbar renormalization scheme
BLOCK BSMINPUTS #
     1      1.00000000E-01   # alpha
     2      0.00000000E+00   # lambda_{phi S}
     3      1.00000000E-01   # lambda_S
     4      0.00000000E+00   # mu_S (GeV)
     5      5.00000000E+02   # mu_r (GeV)
BLOCK SMINPUTS #
     1      7.29735257E-03   # alpha_em
     2      1.16637880E-05   # Fermi constant
     3      1.17900000E-01   # alpha_s
     4      1.27000000E+00   # mc(mc) MSbar
     5      4.18000000E+00   # mb(mb) MSbar
     6      1.67000000E+00   # mc On-shell
     7      4.78000000E+00   # mb On-shell
BLOCK MASS #
     4      6.79114206E-01   # mc(mh) MSbar
     5      2.85835750E+00   # mb(mh) MSbar
     6      1.72500000E+02   # mt
    13      1.05658375E-01   # mmu
    15      1.77686000E+00   # mtau
    23      9.11876000E+01   # mz
    24      8.09388642E+01   # mw (calculated,tree)
    24      8.04085771E+01   # mw (calculated,1-loop)
    25      1.25250000E+02   # mh
    35      5.00000000E+02   # mH
BLOCK CONSTRAINTS #
     0      3.00000000E+03   # The cutoff scale (GeV)
     1      0   # Vacuum stability at tree level [0=OK, 1=No]
     2      0   # Tree-level unitarity [0=OK, 1=No]
     3      0   # S and T parameters [0=OK, 1=No]
     4      0   # True vacuum [0=OK, 1=No]
     5      0   # Vacuum stability (RGE improved with the cutoff scale) [0=OK, 1=No]
     6      0   # Triviality (with the cutoff scale) [0=OK, 1=No]
#
# Partial decay widths of the SM-like Higgs boson by H-COUP #
#          PDG          Width
DECAY       25     0.40743346E-02   # EW:NLO QCD:NNLO                
#          Gamma      NDA       ID1      ID2
     1.42029147E-04     2         4       -4    # Gamma(h -> c c~)
     2.45364374E-03     2         5       -5    # Gamma(h -> b b~)
     8.82215414E-07     2        13      -13    # Gamma(h -> mu- mu+)
     2.54256192E-04     2        15      -15    # Gamma(h -> tau- tau+)
     3.32918565E-04     2        21       21    # Gamma(h -> g g)
     9.25841223E-06     2        22       22    # Gamma(h -> gam gam)
     6.47474306E-06     2        22       23    # Gamma(h -> gam Z)
     9.00793124E-05     2        23       23    # Gamma(h -> Z Z*)
     7.84792254E-04     2        24      -24    # Gamma(h -> W+ W-*)
#
# Partial decay widths of the additional CP-even Higgs boson bH by H-COUP #
#          PDG          Width
DECAY       35     0.87367479E+00   # EW:NLO QCD:NNLO                
#          Gamma      NDA       ID1      ID2
     1.11119396E-01     2         6       -6    # Gamma(bH -> t t~)
     7.77390858E-05     2         5       -5    # Gamma(bH -> b b~)
     4.42785519E-06     2         4       -4    # Gamma(bH -> c c~)
     3.52102715E-08     2        13      -13    # Gamma(bH -> mu- mu+)
     1.01600689E-05     2        15      -15    # Gamma(bH -> tau- tau+)
     1.72174523E-01     2        23       23    # Gamma(bH -> Z Z)
     3.62420028E-01     2        24      -24    # Gamma(bH -> W+ W-)
     2.27571695E-01     2        25       25    # Gamma(bH -> h h)
     2.91017661E-04     2        21       21    # Gamma(bH -> g g)
     2.62532767E-07     2        22       22    # Gamma(bH -> gam gam)
     5.50566818E-06     2        22       23    # Gamma(bH -> gam Z)
\end{lstlisting}

\begin{lstlisting}[caption=Example of the output file (outBR\_hsm.txt),label=file_outBR,frame=lines]
#======================================================================
#
#               H-COUP [Version 3.0 (Novemver 28, 2023)]
#      Program for full-NLO predictions of any Higgs-boson decays
#                     in non-minimal Higgs models
#
#             http://www-het.phys.sci.osaka-u.ac.jp/~hcoup
#
#======================================================================
BLOCK MODEL #
     1      1   # HSM
     2      0   # MSbar renormalization scheme
BLOCK BSMINPUTS #
     1      1.00000000E-01   # alpha
     2      0.00000000E+00   # lambda_{phi S}
     3      1.00000000E-01   # lambda_S
     4      0.00000000E+00   # mu_S (GeV)
     5      5.00000000E+02   # mu_r (GeV)
BLOCK SMINPUTS #
     1      7.29735257E-03   # alpha_em
     2      1.16637880E-05   # Fermi constant
     3      1.17900000E-01   # alpha_s
     4      1.27000000E+00   # mc(mc) MSbar
     5      4.18000000E+00   # mb(mb) MSbar
     6      1.67000000E+00   # mc On-shell
     7      4.78000000E+00   # mb On-shell
BLOCK MASS #
     4      6.79114206E-01   # mc(mh) MSbar
     5      2.85835750E+00   # mb(mh) MSbar
     6      1.72500000E+02   # mt
    13      1.05658375E-01   # mmu
    15      1.77686000E+00   # mtau
    23      9.11876000E+01   # mz
    24      8.09388642E+01   # mw (calculated,tree)
    24      8.04085771E+01   # mw (calculated,1-loop)
    25      1.25250000E+02   # mh
    35      5.00000000E+02   # mH
BLOCK CONSTRAINTS #
     0      3.00000000E+03   # The cutoff scale (GeV)
     1      0   # Vacuum stability at tree level [0=OK, 1=No]
     2      0   # Tree-level unitarity [0=OK, 1=No]
     3      0   # S and T parameters [0=OK, 1=No]
     4      0   # True vacuum [0=OK, 1=No]
     5      0   # Vacuum stability (RGE improved with the cutoff scale) [0=OK, 1=No]
     6      0   # Triviality (with the cutoff scale) [0=OK, 1=No]
#
# Decay branching ratios of the SM-like Higgs boson by H-COUP #
#          PDG          Width
DECAY       25     0.40743346E-02   # EW:NLO QCD:NNLO                
#          BR         NDA       ID1      ID2
     3.48594707E-02     2         4       -4    # BR(h -> c c~)
     6.02219501E-01     2         5       -5    # BR(h -> b b~)
     2.16529938E-04     2        13      -13    # BR(h -> mu- mu+)
     6.24043478E-02     2        15      -15    # BR(h -> tau- tau+)
     8.17111502E-02     2        21       21    # BR(h -> g g)
     2.27237407E-03     2        22       22    # BR(h -> gam gam)
     1.58915350E-03     2        22       23    # BR(h -> gam Z)
     2.21089630E-02     2        23       23    # BR(h -> Z Z*)
     1.92618509E-01     2        24      -24    # BR(h -> W+ W-*)
#
# Decay branching ratios of the additional CP-even Higgs boson bH by H-COUP #
#          PDG          Width
DECAY       35     0.87367479E+00   # EW:NLO QCD:NNLO                
#          BR         NDA       ID1      ID2
     1.27186222E-01     2         6       -6    # BR(bH -> t t~)
     8.89794312E-05     2         5       -5    # BR(bH -> b b~)
     5.06808167E-06     2         4       -4    # BR(bH -> c c~)
     4.03013477E-08     2        13      -13    # BR(bH -> mu- mu+)
     1.16291199E-05     2        15      -15    # BR(bH -> tau- tau+)
     1.97069350E-01     2        23       23    # BR(bH -> Z Z)
     4.14822577E-01     2        24      -24    # BR(bH -> W+ W-)
     2.60476436E-01     2        25       25    # BR(bH -> h h)
     3.33096095E-04     2        21       21    # BR(bH -> g g)
     3.00492552E-07     2        22       22    # BR(bH -> gam gam)
     6.30173635E-06     2        22       23    # BR(bH -> gam Z)
\end{lstlisting}

\section{Examples of numerical evaluations} \label{sec:Discussion}

As mentioned in Introduction, 
it is quite important to include radiative corrections to the analyses 
for the synergy between the precise measurements of $h$ and the direct searches for additional Higgs bosons in order to test and discriminate the extended Higgs models.
We here show an example of such analyses by using \verb|H-COUP_3.0|. 

\begin{figure}
 \centering
 \includegraphics[width=80mm]{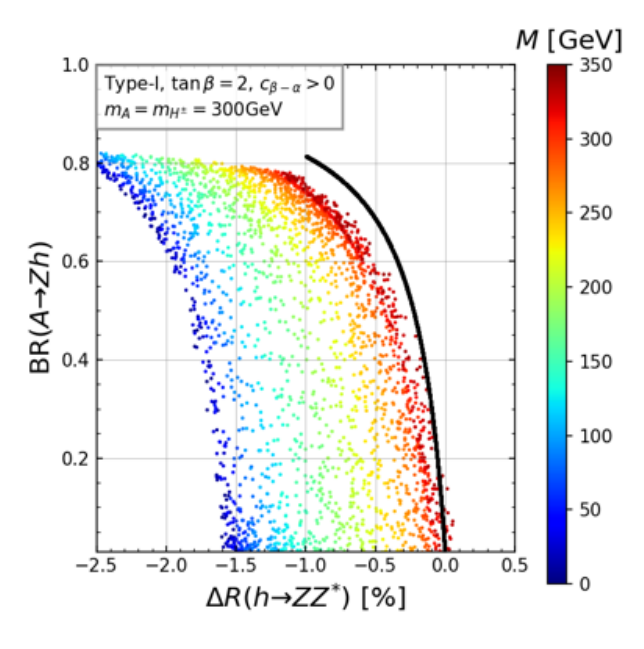}\vspace{0cm}
  \includegraphics[width=80mm]{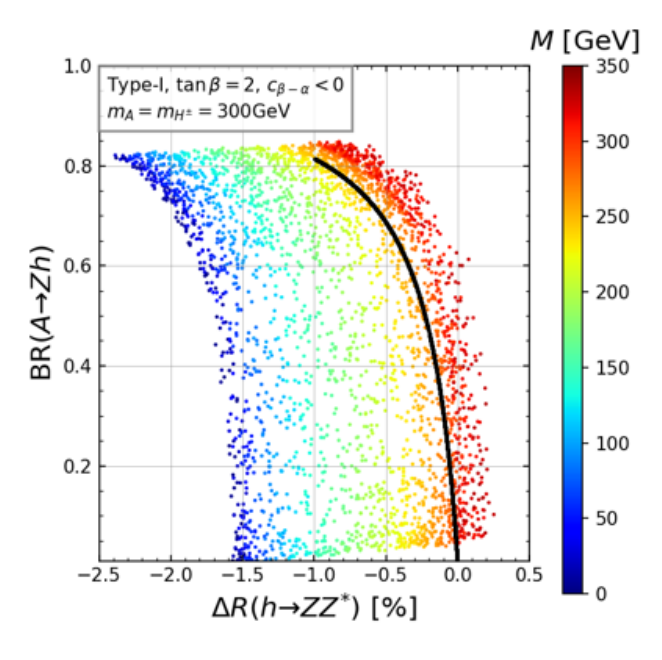}\vspace{0cm}
 \caption{The correlation between decay branching ratios of the $A\to Zh$ process in the Type-I THDM and the deviation of the decay rate of $h\to ZZ$ in the Type-I THDM from that in the SM~\cite{Aiko:2022gmz}. }  
  \label{fig:corr}
\end{figure}

In Fig.~\ref{fig:corr}, we show the correlation between the decay branching ratio of the $A\to Zh$ process  
and the deviation of the decay rate of $h\to ZZ^*$ from the SM prediction in the Type-I THDM.  We define $\Delta R(h\to ZZ^*)\equiv \Gamma[h\to ZZ^*]/\Gamma_\textrm{SM}[h\to ZZ^*]$ to parameterize the deviation~\cite{Aiko:2022gmz}. 
We take $m_A = m_H=300$ GeV and $\tan\beta = 2$, while the values of $\sin(\beta-\alpha)$ and $M^2$ are scanned under the constraints of the perturbative unitarity, the vacuum stability and the electroweak $S$ and $T$ parameters.
We also take into account the constraints from flavor measurements, direct search results of additional Higgs bosons and Higgs coupling measurements.
The left panel and the right panel show results with $\cos(\beta-\alpha) < 0$ and 
$\cos(\beta-\alpha) > 0$, respectively. 
Black curves represent tree-level predictions, while the color dots show the results including radiative corrections with different colors denoting those given by different values of $M^2$. 
It is clearly seen that the radiative corrections significantly change the 
correlation predicted at LO. 
In particular, the case with smaller values of $M^2$ show the larger difference between the results at LO and NLO, because of the larger non-decoupling loop effects of the additional Higgs bosons.
%We note that the size of the radiative corrections to the branching ratio of $A \to Zh$ is typically  
%With this parameter set, the branching ratio of $A\to Zh$ can receive the $\mathcal{O}$(10) \% correction. 
%In this figure, the correlation between the branching ratio of $A\to Zh$ and $\Delta R(h\to ZZ^*)$ is significantly changed from the LO prediction, which indicates the need for both precision measurements of $h$ and direct searches, and the importance of precise calculations for both decays of $h$ and those of additional Higgs bosons. 

In addition to this particular example, it has been known that radiative corrections can significantly 
change the other correlations, e.g., the $H \to hh$ decay rate and the triple Higgs boson coupling $hhh$~\cite{Kanemura:2022ldq}, which can also be evaluated by using \verb|H-COUP_3.0|. 
Therefore, \verb|H-COUP_3.0| can realize the synergy analysis discussed above 
in a more robust way, by which we can reconstruct the structure of the Higgs sector.

\section{Summary} \label{sec:summary}

We have provided the manual for the \verb|H-COUP_3.0| program which is a set of Fortran codes for numerical calculations of decay rates of all the Higgs bosons 
at NLO in electroweak interactions with QCD corrections  
in the HSM, four types of THDMs and the IDM. 
The electroweak corrections are evaluated based on the gauge independent on-shell renormalization scheme. 
%we have added calculations of various decay rates and branching ratios of additional Higgs bosons. 
We have described the renormalization schemes, the renormalized vertex functions and the radiatively corrected decay rates for the processes $\phi\to \phi'\phi''$, $\phi\to V\phi'$ $\phi\to VV'$ and $\phi\to f\bar{f'}$, which are implemented in \verb|H-COUP_3.0|. 
We also have discussed the loop induced decays; i.e. $\phi\to V\gamma$, $\phi\to gg$ and $\phi\to W^\pm Z$, which can be evaluated at LO in electroweak interactions in \verb|H-COUP_3.0|. 
%
%loop corrected decay rates for a set of decay types for scalar bosons; i.e. $\phi\to \phi'\phi''$, $\phi\to V\phi'$ $\phi\to VV'$ and $\phi\to f\bar{f'}$, and decay rates of loop induced processes; i.e. $\phi\to V\gamma$, $\phi\to gg$ and $\phi\to W^\pm Z$.
We have designed the program in a way that the user can choose 
the different two (four) renormalization schemes for the tadpole in the HSM (THDMs), among which the decay rate of $\phi\to \phi'\phi''$ can be different.  
We then have explained the structure of the program, and have shown the examples of the input and output files. 
Finally, we have demonstrated the analysis by using \verb|H-COUP_3.0|. 
As an example, we have shown the loop-corrected correlation between the branching ratio of $A \to Zh$ and the deviation in the $h \to ZZ^*$ decay in the Type-I THDM.  
%\verb|H-COUP_3.0| can realize the to determine the structure of the Higgs sector through the synergy of precise measurement of $h$ and direct search for additional Higgs bosons.

%In this paper, the concept and the manual of {\tt H-COUP\_2.0} have been presented, 
%which is a set of fortran programs for numerical evaluation of decay rates of the Higgs boson %with a mass of 125 GeV  
%and the decay width with higher order corrections (NLO for EW and scalar loop corrections, and %NNLO for QCD corrections) for  
%various models of extended Higgs sectors. 
%In {\tt H-COUP\_2.0}, in addition to the SM, 
%the Higgs singlet model, four types of two Higgs doublet models with a softly-broken $Z_2$ %symmetry and the inert doublet model are implemented. 
%{\tt H-COUP\_2.0} contains all the functions of {\tt H-COUP\_1.0} where a full set of the Higgs %boson vertices are evaluated at one-loop level 
%in a gauge invariant manner in these models. 
%We have briefly introduced these models with their theoretical and experimental constraints, %and we have summarized formulae for the renormalized vertices and the decay rates. 
%After the explanation of the structure of the program, we have demonstrated how to install and %run {\tt H-COUP\_2.0} with some numerical examples. 
%

\begin{acknowledgments}
This work is supported in part by the Grant-in-Aid on Innovative Areas, the Ministry of Education, Culture, 
Sports, Science and Technology, No.~20H00160 and No.~23K17691 [S.K.],
JSPS KAKENHI Grant No.~23KJ0086 and the National Science Centre, Poland, under research Grant No. 2020/38/E/ST2/00243 [K.S.], 
Early-Career Scientists, No.~20K14474 [M.K.] and  
JSPS KAKENHI Grant No.~22KJ3126 [M.A.].
\end{acknowledgments}

\appendix

\section{Scheme difference for \texorpdfstring{$\overline{\rm MS}$}{} parameters in THDMs} \label{app: schemediff}
We calculate the scheme difference for $hhh$, $Hhh$, $HAA$, and $HH^+H^-$ vertices in THDMs. 
We focus on the four schemes presented in Sec.~\ref{sec: reno}. 
The difference between two schemes is defined by
\begin{align}
  \Delta \Gamma(\mbox{ABC}-\mbox{A$^\prime$B$^\prime$C$^\prime$})\equiv
  \Gamma(\mbox{ABC})-\Gamma(\mbox{A$^\prime$B$^\prime$C$^\prime$}).
\end{align}
In Appendix, the trigonometric functions $\sin\theta$ and $\cos\theta$ are represented using the shorthand notation as $s_\theta$ and $c_\theta$. 

As discussed in the main text, the counterterms $\delta M^2$ and $\delta m^2_{12}$ are defined in the  $\overline{\rm MS}$ scheme. 
Analytical expressions for these counterterms are derived in the following way. The counterterms $\delta M^2$ or $\delta m_{12}^2$ absorb the remnant of the UV divergence of the renormalized $hhh$ vertex without these counterterms. 
In the KOSY1 scheme, this condition yields~\cite{Kanemura:2015mxa}
\begin{align}
\left.\frac{\delta M^2}{M^2}\right|_{\rm KOSY1}
 &= \frac{1}{16\pi^2 v^2}
\Big[2\sum_f N_c^f m_f^2 \zeta_f^2 +4M^2 -2m_{H^\pm}^2 -m_A^2  \\ \notag
&+ \frac{s_{2\alpha}^{}}{s_{2\beta}^{}}(m_H^2-m_h^2)-3(2m_W^2+m_Z^2)\Big]\Delta_\textrm{div},
\end{align}
where $\Delta_{\rm div}$ represents the divergent part given by $\Delta_\textrm{div} = 1/\epsilon -\gamma_E + \log 4\pi$. 
Definitions and the explicit expression of coupling factor $\zeta_f$ are given in Ref.~\cite{Kanemura:2022ldq}. 

In the PT1 scheme, additional UV divergent contributions in the scheme difference between KOSY1 and PT1 are absorbed by $\delta M^2$~\cite{Kanemura:2017wtm},
\begin{align} \label{eq:delMPT1}
  \left.\frac{\delta M^2}{M^2}\right|_{\rm PT1}=
  \left.\frac{\delta M^2}{M^2}\right|_{\rm KOSY1}
  +\frac{2}{v}\frac{c_{2\beta}}{s_{2\beta}}
  \left(\frac{\Gamma^{\rm 1PI}_h}{m_h^2}c_{\beta-\alpha}-\frac{\Gamma^{\rm 1PI}_H}{m_H^2}s_{\beta-\alpha}\right)_{\rm div.} \;,
\end{align}
where ``div.'' denotes the UV divergent part. 
As shown in Ref.~\cite{Kanemura:2017wtm}, the last term corresponds to the scheme difference between the KOSY scheme and the PT scheme in $\delta \beta$,
\begin{align}
  \Delta \delta \beta (\mbox{PT}-\mbox{KOSY})=  
  -\frac{1}{v}
  \left(\frac{\Gamma^{\rm 1PI}_h}{m_h^2}c_{\beta-\alpha}-\frac{\Gamma^{\rm 1PI}_H}{m_H^2}s_{\beta-\alpha}\right)\;. 
\end{align}
Thus, Eq.~\eqref{eq:delMPT1} can be rewritten by
\begin{align}
  \left.\frac{\delta M^2}{M^2}\right|_{\rm PT1}=
  \left.\frac{\delta M^2}{M^2}\right|_{\rm KOSY1}
  -{2}\frac{c_{2\beta}}{s_{2\beta}}
  \Delta \delta \beta (\mbox{PT}-\mbox{KOSY})_{\rm div.}
   \;.
\end{align}

From the definition of $M^2$, i.e., $M^2=m^2_{12}/(c_\beta s_\beta)$, the counterterm $\delta m_{12}^2$ in the KOSY2 scheme and the PT2 scheme are obtained from the KOSY1 scheme and the PT1 scheme by the replacement,
\begin{align}
  \delta M_{}^2\to M_{}^2\frac{\delta m_{12}^2}{m_{12}^2}-2{M^2}\frac{c_{2\beta}}{s_{2\beta}}{\overline{\delta \beta}}, 
\end{align} 
with $\overline{\delta \beta}={\delta \beta}+{\delta \beta}^{\rm PT}$ where ${\delta \beta}^{\rm PT}$ is the pinch term contribution for $\delta \beta$. 
Requiring cancellation of the UV divergence in the $hhh$ vertex, in the KOSY2 and PT2 schemes, one obtains 
\begin{align}
  \left.\frac{\delta m^2_{12}}{m^2_{12}}\right|_{\rm KOSY2, PT2} &=
  \left.\frac{\delta M^2}{M^2}\right|_{\rm KOSY1}
  +2\frac{c_{2\beta}}{s_{2\beta}}\left.\delta \beta \right|_{\rm div.}\;.
\end{align}
We note that  $\delta m^2_{12}$ is common between KOSY2 and PT2 since  there is no scheme difference  (see e.g., Eq.~\eqref{eq:difhhh2}.). 
Also, note that there is no UV divergence in $\delta \beta^{\rm PT}$. 
The differences among these schemes are evaluated as
\begin{align}
  \label{eq:difhhh}\Delta \hat{\Gamma}_{hhh}(\mbox{PT1}-\mbox{KOSY1})
  &=
  -\frac{12M^2}{{v^{2}}}  \frac{c_{2\beta} c_{\alpha+ \beta}c_{\beta-\alpha}^2}{s^2_{2 \beta}}   \left(\frac{{\Gamma}_h^{\rm 1PI}}{m_h^2}  c_{\beta - \alpha}-\frac{{\Gamma}_H^{\rm 1PI}}{m_H^2}  s_{\beta - \alpha}\right)_{\rm fin.}\;, \\ \label{eq:difhhh2}
  \Delta \hat{\Gamma}_{hhh}(\mbox{PT2}-\mbox{KOSY2})&=0\;, \\ \label{eq:difbhhh}\
    \Delta \hat{\Gamma}_{Hhh}(\mbox{PT1}-\mbox{KOSY1})
  &=
  -\frac{4M^2}{{v^{2}}}  \frac{c_{2\beta} c_{\beta-\alpha}}{s^2_{2\beta}}(3s_\alpha c_\alpha-s_\beta c_\beta)  \notag \\
&\times  \left(\frac{{\Gamma}_h^{\rm 1PI}}{m_h^2}  c_{\beta - \alpha}-\frac{{\Gamma}_H^{\rm 1PI}}{m_H^2}  s_{\beta - \alpha}\right)_{\rm fin.}\;, \\
  \Delta \hat{\Gamma}_{Hhh}(\mbox{PT2}-\mbox{KOSY2})&=0\;, \\ \label{eq:difbhAA}\
  \Delta \hat{\Gamma}_{HAA}(\mbox{PT1}-\mbox{KOSY1})
  &=
  -\frac{M^2}{{v^{2}}}  \frac{ c_{2\beta}}{ s^2_{\beta}} \frac{s_{\alpha+ \beta}}{c^2_{\beta}}   \left(\frac{{\Gamma}_h^{\rm 1PI}}{m_h^2}  c_{\beta - \alpha}-\frac{{\Gamma}_H^{\rm 1PI}}{m_H^2}  s_{\beta - \alpha}\right)_{\rm fin.}\;, \\
  \Delta \hat{\Gamma}_{HAA}(\mbox{PT2}-\mbox{KOSY2})&=0\;, \\ 
  \Delta \hat{\Gamma}_{HH^+H^-}(\mbox{PT1}-\mbox{KOSY1})&=\Delta \hat{\Gamma}_{HAA}(\mbox{PT1}-\mbox{KOSY1})\;, \label{eq:difHHpHm}\\
  \Delta \hat{\Gamma}_{HH^+H^-}(\mbox{PT2}-\mbox{KOSY2})&=0\;,
\end{align}
where ``fin.'' denotes the finite part.
%We note that one holds $\Delta \delta C_{H^\pm}=0$. Thus, the result for the $HH^+H^-$ vertex can be derived from that of the $HAA$ vertex.
We note that Eq.~\eqref{eq:difHHpHm} is obtained from the fact that one holds $\Delta \delta C_{H^\pm}(\mbox{PT1}-\mbox{KOSY1})=0$.

%%%%%%%%%%%%%%%%%%%%%%%%%%%%%%%%%%%%%%%%%%%%%%%%%%%%%%%%%%%%%%%%%%
\section{Renormalized \texorpdfstring{$\phi\phi'\phi''$}{} vertex} \label{app: phiphiphi} 
In this section, we give formulae of $\Gamma_{\phi\phi'\phi''}^\textrm{loop}$ of several scalar three-point vertices in the THDM and the IDM, which is defined in Eq.~(\ref{eq:form-loop}) in Sec.~\ref{sec: reno}. 

%$\Gamma_{\phi\phi'\phi''}^\textrm{loop}$ is further divided into contributions from counter terms and those from 1PI diagrams, i.e. $\delta\Gamma_{\phi\phi'\phi''}$ and $\Gamma_{\phi\phi'\phi''}^\textrm{1PI}$, respectively. 

\subsection{\texorpdfstring{$HAA$}{} and \texorpdfstring{$HH^+H^-$}{} vertices in the THDM} 
First, we show formulae for the $HAA$ and $HH^+H^-$ vertices in the THDM. 
We here give the formulas for the KOSY1 scheme defined in Sec.~\ref{sec: reno}. 
Conversions to other schemes are explained in the previous section. 

Counterterms for the $HAA$ and the $HH^+H^-$ vertices in the THDM are expressed as 
\begin{align}
  \delta\Gamma_{HAA}^{} &=2\delta\lambda_{HAA}^{} + 2\lambda_{HAA}^{}(\delta Z_A +\frac{1}{2}\delta Z_H) + 2\lambda_{hAA}(\delta C_{h} -\delta\alpha)
  +2\lambda_{HAG}^{}(\delta C_{A} +\delta\beta), \\
  \delta\Gamma_{HH^+H^-}^{} &=\delta\lambda_{HH^+H^-}^{} + \lambda_{HH^+H^-}^{}(\delta Z_{H^+} +\frac{1}{2}\delta Z_H) + \lambda_{hH^+H^-}(\delta C_{h} -\delta\alpha)\notag\\
  &+2\lambda_{HH^+G^-}^{}(\delta C_{H^{\pm}} +\delta\beta), 
\end{align}
with
\begin{align}
  \delta\lambda_{HAA}^{} &=
  -\frac{\lambda_{HAA}^{}}{v}\delta v
  - \frac{c_{\beta-\alpha}^{}}{v}\delta m_A^2
  -\frac{s_{\alpha -3\beta}+3s_{\alpha +\beta}}{8vs_\beta^{}c_\beta^{}}\delta m_H^2
  + \frac{s_{\alpha +\beta}}{2vs_{\beta}^{}c_\beta^{}}\delta M^2 
  +G_\alpha^A\overline{\delta\alpha} 
  +G_\beta^A\overline{\delta\beta}, \label{eq:deltaHAA}\\
  % &\delta\lambda_{HH^+H^-}^{} = 2\delta\lambda_{HAA}^{} \quad ( m_A^2 \to m_{H^\pm}^2, \, \delta m_A^2 \to \delta m_{H^\pm}^2).   
  \delta\lambda_{HH^+H^-}^{} &= 2\delta\lambda_{HAA}^{} \quad (A \to H^{\pm}).   
  \end{align}
In Eq.~(\ref{eq:deltaHAA}), the factors $G_\alpha^\phi$ and $G_\beta^\phi$ are expressed as
\begin{align}
  &G_\alpha^\phi =-\frac{1}{8vs_\beta^{}c_\beta^{}}\left[
  c_{\alpha-3\beta}^{}m_H^2 + c_{\alpha +\beta}^{}(3m_H^2-4M^2) + 8s_{\beta-\alpha}^{}s_\beta^{}c_\beta^{}m_\phi^2 
  \right],\\
  &G_\beta^\phi =\frac{1}{8vs_{2\beta}^2}\left[
  (2m_\phi^2 -m_H^2)s_{\alpha-5\beta}^{} + 2(2m_\phi^2 -7m_H^2 +6M^2)s_{\beta-\alpha}^{}
  +(2m_\phi^2 +3m_H^2 -4M^2)s_{\alpha+3\beta}^{}
  \right]. 
\end{align}
%where $c_{A}$ $(c_{H^+}^{})=1$, $2$. 

The analytic expressions for 1PI diagram contributions to the $HAA$ and $HH^+H^-$ vertices are given by
\begin{align}
 & \left(16\pi^2\right)\Gamma_{HAA, F}^\textrm{1PI}[p_1^2,p_2^2,q^2] = -8\kappa_{f}^{H}\zeta_f^2 N_c^f\frac{m_f^4}{v^3}\left(
  B_0[q^2;m_f,m_f] + p_1\cdot p_2 C_0[m_f,m_f,m_f]\right),  \label{eq:HAA_F}\\
  \notag\\
  & \left(16\pi^2\right)\Gamma_{HH^+H^-, F}^\textrm{1PI}[p_1^2,p_2^2,q^2] =\notag\\
  &-\frac{4m_f^2}{v^3}N_c^f\kappa_f^H\Bigg\{
  2(m_f^2\zeta_f^2 + m_{f'}^2 \zeta_{f'}^2 - m_{f'}^2 \zeta_f \zeta_{f'})\left(p_1^2 C_{21}^{} + p_2^2C_{22}+2p_1\cdot p_2 C_{23} +4C_{24} -\frac{1}{2}\right)
  \notag\\
  &+(m_f^2 \zeta_f^2 +m_{f'}^2 \zeta_{f'}^2)\left[(2p_1^2 +p_1\cdot q)C_{11} +(2p_1\cdot p_2 +p_2\cdot q)C_{12} +p_1\cdot q C_0 \right]\notag\\
  & -2m_{f'}^2 \zeta_f \zeta_{f'}(p_1\cdot q C_{11} +p_2\cdot q C_{12} +m_f^2 C_0)\Bigg\}[f,f',f],\label{eq:HHpHm_F}\\ 
  \notag\\
%\end{align}
%\begin{align}
  &\left(16\pi^2\right)\Gamma_{HAA, B}^\textrm{1PI}[p_1^2,p_2^2,q^2]= \notag\\
  &\frac{g^4}{4}vc_{\beta-\alpha}^{}(4B_0[q^2;W,W]-2)
  +\frac{g_Z^4}{8}vc_{\beta-\alpha}^{}(4B_0[q^2;Z,Z]-2) \notag\\
  &-\frac{g^4}{4}vc_{\beta-\alpha}^{}C_{VSV}^{\phi\phi\phi}[W,H^\pm,W]
  -\frac{g_Z^4}{8}vc_{\beta-\alpha}^{3}C_{VSV}^{\phi\phi\phi}[Z,h,Z]
  -\frac{g_Z^4}{8}vs_{\beta-\alpha}^2c_{\beta-\alpha}^{}C_{VSV}^{\phi\phi\phi}[Z,H,Z]  \notag\\
  &-\frac{g^2}{2}c_{\beta-\alpha}^{}\frac{m_A^2-m_{H^\pm}^2}{v}C_{VSS}^{\phi\phi\phi}[W,H^\pm,G^\pm] 
  +\frac{g^2}{2}\lambda_{HH^+H^-}^{}C_{SVS}^{\phi\phi\phi}[H^\pm,W,H^\pm] \notag\\
  &-\frac{g^2}{2}c_{\beta-\alpha}^{}\frac{m_A^2-m_{H^\pm}^2}{v}C_{SSV}^{\phi\phi\phi}[G^\pm,H^\pm,W] 
   -\frac{g_Z^2}{4}c_{\beta-\alpha}^{2}\lambda_{hAG}^{}C_{VSS}^{\phi\phi\phi}[Z,h,G] \notag\\
  &+\frac{g_Z^2}{4}s_{\beta-\alpha}c_{\beta-\alpha}\lambda_{HAG}^{}C_{VSS}^{\phi\phi\phi}[Z,H,G]
  +\frac{g_Z^2}{2}s_{\beta-\alpha}c_{\beta-\alpha}\lambda_{hAA}^{}C_{VSS}^{\phi\phi\phi}[Z,h,A]
  -\frac{g_Z^2}{2}s_{\beta-\alpha}^2\lambda_{HAA}^{}C_{VSS}^{\phi\phi\phi}[Z,H,A]  \notag\\
  &+\frac{g_Z^2}{2}c_{\beta-\alpha}^{2}\lambda_{Hhh}^{}C_{SVS}^{\phi\phi\phi}[h,Z,h]
  -\frac{g_Z^2}{2}s_{\beta-\alpha}^{}c_{\beta-\alpha}^{}\lambda_{HHh}^{}\{ C_{SVS}^{\phi\phi\phi}[h,Z,H] + C_{SVS}^{\phi\phi\phi}[H,Z,h]\}\notag\\
  &+\frac{3g_Z^2}{2}s_{\beta-\alpha}^{2}\lambda_{HHH} C_{SVS}^{\phi\phi\phi}[H,Z,H]
  -\frac{g_Z^2}{4}c_{\beta-\alpha}^{2}\lambda_{hAG} C_{SSV}^{\phi\phi\phi}[G,h,Z]
  +\frac{g_Z^2}{4}s_{\beta-\alpha}^{}c_{\beta-\alpha}^{}\lambda_{HAG} C_{SSV}^{\phi\phi\phi}[G,H,Z] \notag\\
  &+\frac{g_Z^2}{2}s_{\beta-\alpha}^{}c_{\beta-\alpha}^{}\lambda_{hAA} C_{SSV}^{\phi\phi\phi}[A,h,Z]
  -\frac{g_Z^2}{2}s_{\beta-\alpha}^{2}\lambda_{HAA} C_{SSV}^{\phi\phi\phi}[A,H,Z] \notag\\
  %%%%%%%%%%%%%%%%%%%%%%%%%
  &+2\lambda_{HH^+H^-}^{}\lambda_{AAH^+H^-}^{}B_0[q^2;H^\pm,H^\pm] +4\lambda_{HH^\pm G^\mp}^{}\lambda_{AAH^\mp G^\pm}^{}B_0[q^2;H^\pm,G^\pm]\notag\\
  &+2\lambda_{AH^\pm G^\mp}^{}\lambda_{AHH^\mp G^\pm}^{}B_0[p_2^2;H^\pm,G^\pm]
  +2\lambda_{AH^\pm G^\mp}^{}\lambda_{AHH^\mp G^\pm}^{}B_0[p_1^2;H^\pm,G^\pm] \notag\\
  &  +2\lambda_{HG^\pm G^\mp}^{}\lambda_{AAG^\pm G^\mp}^{}B_0[q^2;G^\pm,G^\pm]
  +4\lambda_{Hhh}^{}\lambda_{hhAA}^{}B_0[q^2;h,h]
  +4\lambda_{HHh}^{}\lambda_{hHAA}^{}B_0[q^2;h,H] \notag\\
  &+12\lambda_{HHH}^{}\lambda_{HHAA}^{}B_0[q^2;H,H]
  +24\lambda_{HAA}^{}\lambda_{AAAA}^{}B_0[q^2;A,A]
  +6\lambda_{HAG}^{}\lambda_{AAAG}^{}B_0[q^2;A,G] \notag\\
  & +4\lambda_{HGG}^{}\lambda_{AAGG}^{}B_0[q^2;G,G]
  +\lambda_{hAG}^{}\lambda_{hHAG}^{}\{B_0[p_1^2;G,h] +B_0[p_2^2;G,h]\} \notag\\
  &  +4\lambda_{hAA}^{}\lambda_{hHAA}^{}\{B_0[p_1^2;A,h] +B_0[p_2^2;A,h]\}
  +2\lambda_{HAG}^{}\lambda_{HHAG}^{}\{B_0[p_1^2;G,H] +B_0[p_2^2;G,H]\} \notag\\
  &  +8\lambda_{HAA}^{}\lambda_{HHAA}^{}\{B_0[p_1^2;H,A] +B_0[p_2^2;H,A]\} \notag\\
  %%%%%%%%%%%%%%%%%%%%%%%%%%
  &- 2\lambda_{HH^+H^-}|\lambda_{AH^\pm G^\mp}|^2C_0[H^\pm,G^\pm,H^\pm] - 2\lambda_{HG^+G^-}|\lambda_{AH^\pm G^\mp}|^2C_0[G^\pm,H^\pm,G^\pm]\notag\\
  &- 8\lambda_{Hhh}\lambda_{hAA}^2C_0[h,A,h]
  -2\lambda_{Hhh}\lambda_{hGA}^2C_0[h,G,h] 
  - 8\lambda_{hHH}\lambda_{hAA}^2C_0[h,A,H]\notag\\
  &-2\lambda_{hHH}\lambda_{hGA}\lambda_{HGA}C_0[h,G,H] -8\lambda_{hHH}\lambda_{hAA}^2C_0[H,A,h]
  -2\lambda_{hHH}\lambda_{hGA}\lambda_{HGA}C_0[H,G,h] \notag\\
  &-24\lambda_{HHH}\lambda_{HAA}^2 C_0[H,A,H]
  -6\lambda_{HHH}\lambda_{HGA}^2 C_0[H,G,H] -8\lambda_{HAA}\lambda_{hAA}^2 C_0[A,h,A] \notag\\
  &-8\lambda_{HAA}^3C_0[A,H,A] -2\lambda_{HGA}\lambda_{hGA}\lambda_{hAA}\{C_0[G,h,A] +C_0[A,h,G]\} \notag\\
  &-2\lambda_{HGA}^2\lambda_{HAA}\{ C_0[G,H,A] +C_0[A,H,G]\}
  -2\lambda_{HGG}\lambda_{hGA}^2 C_0[G,h,G] - 2\lambda_{HGG}\lambda_{HGA}^2C_0[G,H,G], \label{eq:HAA_B}\\
  \notag\\
%  
%\end{align}
%\begin{align}
  &\left(16\pi^2\right)\Gamma_{HH^+H^-, B}^\textrm{1PI}[p_1^2,p_2^2,q^2] =\notag\\
  &\frac{g^4}{4}vc_{\beta-\alpha}^{}(4B_0[q^2;W,W]-2) 
  + \frac{g_Z^4}{8}c_{2W}^2vc_{\beta-\alpha}^{}(4B_0[q^2;Z,Z]-2)\notag\\
  &-\frac{g^4}{8}vc_{\beta-\alpha}^3 C_{VSV}^{\phi\phi\phi}[W,h,W] -\frac{g^4}{8}vs_{\beta-\alpha}^2c_{\beta-\alpha}^{} C_{VSV}^{\phi\phi\phi}[W,H,W]
  -\frac{g^4}{8}vc_{\beta-\alpha}^{} C_{VSV}^{\phi\phi\phi}[W,A,W]\notag\\
  &-\frac{g_Z^4}{8}vc_{2W}^2 c_{\beta-\alpha}^{} C_{VSV}^{\phi\phi\phi}[Z,H^+,Z]
  +\frac{g^2}{4}s_{\beta-\alpha}^{}c_{\beta-\alpha}^{}\lambda_{hH^+ H^-}^{} C_{VSS}^{\phi\phi\phi}[W,h,H^-]\notag\\
  &-\frac{g^2}{4}s_{\beta-\alpha}^{2}\lambda_{HH^+ H^-}^{} C_{VSS}^{\phi\phi\phi}[W,H,H^-] 
   -\frac{g^2}{4}c_{\beta-\alpha}^{2}\lambda_{hG^+ H^-}^{} C_{VSS}^{\phi\phi\phi}[W,h,G^-]\notag\\
  &+\frac{g^2}{4}s_{\beta-\alpha}^{}c_{\beta-\alpha}^{}\lambda_{HG^+ H^-}^{} C_{VSS}^{\phi\phi\phi}[W,H,G^-]
  -\frac{g^2}{4}c_{\beta-\alpha}^{}\frac{m_{H^+}^2-m_A^2}{v} C_{VSS}^{\phi\phi\phi}[W,A,G^-]\notag\\
  & +\frac{g^2}{2}c_{\beta-\alpha}^{2}\lambda_{hhH}^{} C_{SVS}^{\phi\phi\phi}[h,W,h]
  -\frac{g^2}{2}s_{\beta-\alpha}^{}c_{\beta-\alpha}^{}\lambda_{hHH}^{} \{C_{SVS}^{\phi\phi\phi}[h,W,H] +C_{SVS}^{\phi\phi\phi}[H,W,h]\}\notag\\
  &+\frac{3g^2}{2}s_{\beta-\alpha}^{2}\lambda_{HHH}^{} C_{SVS}^{\phi\phi\phi}[H,W,H]
  +\frac{g^2}{2}\lambda_{HAA}^{} C_{SVS}^{\phi\phi\phi}[A,W,A]
  +\frac{g_Z^2}{4}c_{2W}^2\lambda_{HH^+H^-}^{} C_{SVS}^{\phi\phi\phi}[H^-,Z,H^-]\notag\\
  &+e^2 \lambda_{HH^+H^-}C_{SVS}^{\phi\phi\phi}[H^-,\gamma,H^-]
  +\frac{g^2}{4}s_{\beta-\alpha}^{}c_{\beta-\alpha}^{}\lambda_{hH^+ H^-}^{} C_{SSV}^{\phi\phi\phi}[H^-,h,W] \notag\\
  &-\frac{g^2}{4}s_{\beta-\alpha}^{2}\lambda_{HH^+ H^-}^{} C_{SSV}^{\phi\phi\phi}[H^-,H,W]
  -\frac{g^2}{4}c_{\beta-\alpha}^{2}\lambda_{hH^+ G^-}^{} C_{SSV}^{\phi\phi\phi}[G^-,h,W] \notag\\
  &+\frac{g^2}{4}s_{\beta-\alpha}^{}c_{\beta-\alpha}^{}\lambda_{HH^+ G^-}^{} C_{SSV}^{\phi\phi\phi}[G^-,H,W]
  -\frac{g^2}{4}c_{\beta-\alpha}^{}\frac{m_{H^+}^2-m_A^2}{v} C_{SSV}^{\phi\phi\phi}[G^-,A,W]\notag\\
  %%%
  & +4\lambda_{HH^+H^-}^{}\lambda_{H^+H^-H^+H^-}^{}B_0[q^2;H^+,H^+]
  +\lambda_{HG^+G^-}^{}\lambda_{H^+H^-G^+G^-}^{}B_0[q^2;G^+,G^+] \notag\\
  &+4\lambda_{HH^+G^-}^{}\lambda_{H^+H^-H^-G^+}^{}B_0[q^2;H^+,G^+]
   +\lambda_{hH^+H^-}^{}\lambda_{HhH^+H^-}^{}(B_0[p_1^2;H^+,h] + B_0[p_2^2;H^+,h] )\notag\\
  &+2\lambda_{HH^+H^-}^{}\lambda_{HHH^+H^-}^{}(B_0[p_1^2;H^+,H] + B_0[p_2^2;H^+,H] )\notag\\
  &+ 2\lambda_{HH^+G^-}^{}\lambda_{HHG^+H^-}^{}B_0[p_1^2;G^-,H]+ 2\lambda_{HG^+H^-}^{}\lambda_{HHH^+G^-}^{}B_0[p_2^2;G^+,H] \notag\\
  &+ \lambda_{hH^+G^-}^{}\lambda_{HhG^+H^-}^{}B_0[p_1^2;G^-,h]+ \lambda_{hG^+H^-}^{}\lambda_{HhH^+G^-}^{}B_0[p_2^2;G^+,h] \notag\\
  &+ \lambda_{AH^+G^-}^{}\lambda_{HAG^+H^-}^{}B_0[p_1^2;G^-,A]+ \lambda_{AG^+H^-}^{}\lambda_{HAH^+G^-}^{}B_0[p_2^2;G^+,A] \notag\\
  & + 2\lambda_{hhH}^{}\lambda_{hhH^+H^-}^{}B_0[q^2;h,h]+ 2\lambda_{hHH}^{}\lambda_{HhH^+H^-}^{}B_0[q^2;h,H] 
  + 6\lambda_{HHH}^{}\lambda_{HHH^+H^-}^{}B_0[q^2;H,H]\notag\\
  &+ 2\lambda_{HAA}^{}\lambda_{AAH^+H^-}^{}B_0[q^2;A,A]+ \lambda_{HAG}^{}\lambda_{GAH^+H^-}^{}B_0[q^2;A,G^0]
  + 2\lambda_{HGG}^{}\lambda_{GGH^+H^-}^{}B_0[q^2;G^0,G^0]\notag\\
  %%%%%
  &-\lambda_{HH^+H^-}^{}(\lambda_{hH^+H^-}^2 C_0[H^-,h,H^-] + \lambda_{HH^+H^-}^2 C_0[H^-,H,H^-]) \notag\\
  &-2\lambda_{hhH}^{}\lambda_{hH^+H^-}^2 C_0[h,H^+,h] - 6\lambda_{HHH}^{}\lambda_{HH^+H^-}^2 C_0[H,H^+,H] \notag\\
  &-2\lambda_{hHH}^{}\lambda_{hH^+H^-}^{}\lambda_{HH^+H^-}^{}( C_0[h,H^+,H] + C_0[H,H^+,h]) \notag\\
  &-\lambda_{HH^+G^-}^{}(\lambda_{hH^+H^-}^{}\lambda_{hG^+H^-}^{} C_0[H^-,h,G^-] + \lambda_{HH^+H^-}^{}\lambda_{HG^+H^-}^{} C_0[H^-,H,G^-]) \notag\\
  &-\lambda_{HG^+H^-}^{}(\lambda_{hH^+H^-}^{}\lambda_{hH^+G^-}^{} C_0[G^-,h,H^-] + \lambda_{HH^+H^-}^{}\lambda_{HH^+G^-}^{} C_0[G^-,H,H^-]) \notag\\
  &-2\lambda_{hhH}^{}\lambda_{hH^+G^-}^2 C_0[h,G^+,h] - 6\lambda_{HHH}^{}\lambda_{HH^+G^-}^2 C_0[H,G^+,H] \notag\\
  &-2\lambda_{hHH}^{}\lambda_{hH^+G^-}^{}\lambda_{HH^+G^-}^{}( C_0[h,G^+,H] + C_0[H,G^+,h]) \notag\\
  &- 2\lambda_{HAA}^{}|\lambda_{AH^+G^-}|^2 C_0[A,G^+,A] 
  -\lambda_{HG^+G^-}^{}(\lambda_{hH^+G^-}^2 C_0[G^-,h,G^-] + \lambda_{HH^+G^-}^2 C_0[G^-,H,G^-]) \notag\\
  &-\lambda_{HG^+G^-}^{}|\lambda_{AH^+G^-}|^2 C_0[G^-,A,G^-], \label{eq:HHpHm_B}
\end{align}
where the definition of the coupling factor $\kappa_f^H$ is given in Ref.~\cite{Kanemura:2022ldq}. 
$B_i$ and $C_i$ functions represent Passarino-Veltman functions~\cite{Passarino:1978jh} with
 \begin{align}
 B_i[p_j^2;X,Y]&\equiv B_i[p_j^2;m_X, m_Y],\\
   C_i[X, Y, Z] &\equiv C_i[p_1^2,p_2^2,q^2; m_X, m_Y, m_Z]. 
   \end{align}
Each combination of $C^{}$-functions is defined as, 
\begin{align}
  C_{SVV}^{\phi\phi\phi}[S, V_1, V_2]&= [p_1^2 C_{21}^{} + p_2^2C_{22}+2p_1\cdot p_2 C_{23} +4C_{24} -\frac{1}{2} \notag\\[8pt]
 &- (q+ p_1^{}) \cdot(p_1 C_{11} + p_2 C_{12}) + q\cdot p_1 C_0](S, V_1, V_2), \\
  C_{VSV}^{\phi\phi\phi}[V_2,S, V_1]&= [p_1^2 C_{21}^{} + p_2^2C_{22}+2p_1\cdot p_2 C_{23} +4C_{24} -\frac{1}{2} \notag\\[8pt]
 &+ (3p_1^{} -p_2) \cdot(p_1 C_{11} + p_2 C_{12}) + 2p_1\cdot (p_1 - p_2) C_0](V_2, S, V_1), \\[8pt]
  C_{VVS }^{\phi\phi\phi}[V_1, V_2, S]&= [p_1^2 C_{21}^{} + p_2^2C_{22}+2p_1\cdot p_2 C_{23} +4C_{24} -\frac{1}{2} \notag\\[8pt]
 &+ (3p_1^{} +4p_2) \cdot(p_1 C_{11} + p_2 C_{12}) + 2q\cdot (q +p_2) C_0](V_1, V_2, S), \\[8pt]
  C_{VSS}^{\phi\phi\phi}[V, S_1, S_2]&= [p_1^2 C_{21}^{} + p_2^2C_{22}+2p_1\cdot p_2 C_{23} +4C_{24} -\frac{1}{2} \notag\\[8pt]
  &+ (4p_1^{} +2p_2) \cdot(p_1 C_{11} + p_2 C_{12}) + 4q\cdot p_1  C_0](V, S_1, S_2), \\[8pt]
   C_{SVS}^{\phi\phi\phi}[S_2, V, S_1]& = [p_1^2 C_{21}^{} + p_2^2C_{22}+2p_1\cdot p_2 C_{23} +4C_{24} -\frac{1}{2} \notag\\[8pt]
  &+ 2p_2\cdot(p_1 C_{11} + p_2 C_{12}) -p_1\cdot (p_1 +2 p_2) C_0](S_2, V,S_1), \\[8pt]
  C_{SSV}^{\phi\phi\phi}[S_1, S_2, V]&= [p_1^2 C_{21}^{} + p_2^2C_{22}+2p_1\cdot p_2 C_{23} +4C_{24} -\frac{1}{2} \notag\\[8pt]
   &- 2p_2 \cdot(p_1 C_{11} + p_2 C_{12}) -q \cdot (p_1 - p_2) C_0](S_1, S_2, V). 
 \end{align}
The first two Eqs.~(\ref{eq:HAA_F}) and (\ref{eq:HHpHm_F}) and the latter two Eq.~(\ref{eq:HAA_B}) and (\ref{eq:HHpHm_B}) are fermion loop contributions and boson loop contributions, respectively. 
The definition of momentum is given in Sec.~\ref{sec: reno}.

As mentioned in Sec.~\ref{sec: reno}, the non-vanishing contribution from mixing self-energies is included in the electroweak corrections to the decay rate $H\to H^+H^-$, which is calculated as,
\begin{align}
\Delta_\textrm{mix}(H\to H^+H^-) = 
-4\frac{\lambda_{HH^+G^-}^{}}{\lambda_{HH^+H^-}^{}}
\frac{\hat{\Pi}_{H^+G^-}(m_{H^\pm}^2)}{m_{H^\pm}^2}.
\end{align}

\subsection{\texorpdfstring{$hHH$}{}, \texorpdfstring{$hAA$}{}, \texorpdfstring{$hH^+H^-$}{} vertices in the IDM}
We here give explicit formulae for counterterms and 1PI diagram contributions of the $hHH$, $hAA$, $hH^+H^-$ vertices in the IDM. 

The explicit formula of the counterterm of the $h\phi\phi$ ($\phi = H, A, H^\pm$) is given by 
\begin{align}
  &\delta \Gamma_{h\phi\phi}^{} = 2\delta\lambda_{h\phi\phi} + C_\phi\lambda_{h\phi\phi}\left(\delta Z_\phi + \frac{1}{2}\delta Z_h\right), 
\end{align}
with
\begin{align}
 & \delta\lambda_{h\phi\phi}^{} = -\frac{1}{v}\delta m_\phi^2 + \frac{m_\phi^2 - \mu_2^2}{v^2}\delta v + \frac{1}{v}\delta\mu_2^2, \\
 & C_H = C_A = 2, \quad C_{H^\pm}^{} = 1. 
  \end{align}
$\delta \mu_2^2$ is determined by $\overline{\textrm{MS}}$ scheme as well as $\delta M^2$ in the THDM, so that it is given by 
\begin{align}
  \delta \mu_2^2 =\frac{1}{16\pi^2}\left[-\frac{m_h^2}{2v^2}(m_H^2+m_A^2+2m_{H^\pm}^2)+\frac{\mu_2^2}{v^2}(2m_h^2-6m_W^2-3m_Z^2+3\lambda_2^{}v^2)\right]\Delta_\textrm{div}. 
\end{align}

The 1PI diagram contributions to the $h\Phi\Phi$ vertex are calculated as
\begin{align}
  & (16\pi^2)\Gamma_{hHH, B}^\textrm{1PI}[q^2,p_1^2,p_2^2] =\notag\\
  & 12\lambda_{hhh}\lambda_{hhHH}B_0[q^2;h,h] + 24\lambda_{hHH}\lambda_{HHHH} B_0[q^2;H,H] +4\lambda_{hAA}\lambda_{HHAA}B_0[q^2;A,A] \notag\\
  &+ 4\lambda_{hGG}\lambda_{HHGG}B_0[q^2;G^0,G^0] + 2\lambda_{hH^+H^-}\lambda_{HHH^+H^-}B_0[q^2;H^\pm,H^\pm]
  + 2\lambda_{hG^+G^-}\lambda_{HHG^+G^-}B_0[q^2;G^\pm,G^\pm] \notag\\
  & +8\lambda_{hHH}\lambda_{hhHH}(B_0[p_1^2;h,H] + B_0[p_2^2;h,H])
  +\lambda_{HAG}\lambda_{hHAG} ( B_0[p_1^2;G^0,A] + B_0[p_2^2;G^0,A] ) \notag\\
  & +2 \lambda_{HH^+G^-}\lambda_{hHH^-G^+}(B_0[p_1^2;G^\pm,H^\pm] + B_0[p_2^2;G^\pm,H^\pm]) \notag\\
  & + 2g^3 m_W(B_0[q^2;W,W] -\frac{1}{2}) + g_Z^3 m_Z(B_0[q^2;Z,Z] -\frac{1}{2}) \notag\\
  & -\frac{g_Z^4}{8}v C_{VSV}^{\phi\phi\phi}[Z,A,Z] -\frac{g^4}{4}v C_{VSV}^{\phi\phi\phi}[W,H^\pm,W]
  + \frac{g_Z^2}{2}\lambda_{hAA} C_{SVS}^{\phi\phi\phi}[A,Z,A] + \frac{g^2}{2}\lambda_{hH^+H^-}C_{SVS}^{\phi\phi\phi}[H^\pm,W,H^\pm] \notag\\
  & + \frac{g_Z^2}{4}\lambda_{HAG}(C_{VSS}^{\phi\phi\phi}[Z,A,G^0] + C_{SSV}^{\phi\phi\phi}[G^0,A,Z])
  + \frac{g^2}{2}\lambda_{HH^+G^-}(C_{VSS}^{\phi\phi\phi}[W,H^\pm,G^\pm] + C_{SSV}^{\phi\phi\phi}[G^\pm,H^\pm,W]) \notag\\
  & -24\lambda_{hhh}\lambda_{hHH}^2 C_0[h,H,h] -8\lambda_{hHH}^3 C_0[H,h,H] -2\lambda_{hAA}\lambda_{HAG}^2C_0[A,G^0,A]
  -2\lambda_{hGG}\lambda_{HAG}^2C_0[G^0,A,G^0] \notag\\
  & -2\lambda_{hH^+H^-}|\lambda_{HH^+G^-}|^2C_0[H^\pm,G^\pm,H^\pm] - 2\lambda_{hG^+G^-}|\lambda_{HH^+G^-}|^2C_0[G^\pm,H^\pm,G^\pm],\\
 \notag\\
%\end{align}
% hAA 
%\begin{align}
  & (16\pi^2)\Gamma_{hAA}^\textrm{1PI,B}[q^2,p_1^2,p_2^2] =\notag\\
  & 24\lambda_{hAA}\lambda_{AAAA}B_0[q^2;A,A] + 8\lambda_{hAA}\lambda_{hhAA}(B_0[p_1^2;,h,A] + B_0[p_2^2;h,A])
  + 4\lambda_{hHH}\lambda_{HHAA}B_0[q^2;H,H] \notag\\
  &+ 12\lambda_{hhh}\lambda_{hhAA}B_0[q^2;h,h] + 4\lambda_{hGG}\lambda_{AAGG} B_0[q^2;G^0,G^0]
  +\lambda_{HAG}\lambda_{hHAG}(B_0[p_1^2;H,G^0] +B_0[p_2^2;H,G^0]) \notag\\
  & +2\lambda_{hG^+G^-}\lambda_{AAG^+G^-}B_0[q^2;G^\pm,G^\pm] + 2\lambda_{AH^-G^+}\lambda_{hAH^+G^-}(B_0[p_1^2;H^\pm,G^\pm] +B_0[p_2^2;H^\pm,G^\pm]) \notag\\
  & + 2\lambda_{hH^+H^-}\lambda_{AAH^+H^-} B_0[q^2;H^\pm,H^\pm] \notag\\
  & +2g^3 m_W(B_0[q^2;W,W] - \frac{1}{2}) +g_Z^3 m_Z(B_0[q^2;Z,Z] - \frac{1}{2}) \notag\\
  & -\frac{g_Z^4}{8}v C_{VSV}^{\phi\phi\phi}[Z,H,Z]- \frac{g^4}{4}v C_{VSV}^{\phi\phi\phi}[W,H^\pm,W]
  + \frac{g_Z^2}{2}\lambda_{hHH}C_{SVS}^{\phi\phi\phi}[H,Z,H]+ \frac{g^2}{2}\lambda_{hH^+H^-}C_{SVS}^{\phi\phi\phi}[H^\pm,W.H^\pm]\notag\\
  &+i\frac{g^2}{2}\lambda_{AH^+G^-}(C_{VSS}^{\phi\phi\phi}[W,H^\pm,G^\pm] + C_{SSV}^{\phi\phi\phi}[G^\pm,H^\pm,W])
  -\frac{g_Z^2}{4}\lambda_{HAG}(C_{VSS}^{\phi\phi\phi}[Z,H,G^0] + C_{SSV}^{\phi\phi\phi}[G^0,H,Z]) \notag\\
  &-24\lambda_{hhh}\lambda_{hAA}^2C_0[h,A,h] -2\lambda_{hHH}\lambda_{HAG}^2C_0[H,G^0,H] -8\lambda_{hAA}^3C_0[A,h,A]\notag\\
  &-2\lambda_{hGG}\lambda_{HAG}^2C_0[G^0,H,G^0] -2\lambda_{hH^+H^-}\lambda_{AH^+G^-}\lambda_{AH^-G^+}C_0[H^\pm,G^\pm,H^\pm] \notag\\
  &-2\lambda_{hG^+G^-}\lambda_{AH^+G^-}\lambda_{AH^-G^+}C_0[G^\pm,H^\pm,G^\pm], \\
%\end{align}
%\begin{align}
  & (16\pi^2)\Gamma_{hH^+H^-}^\textrm{1PI,B}[q^2,p_1^2,p_2^2] \notag\\
  &= 6\lambda_{hhh}\lambda_{hhH^+H^-} B_0[q^2;h,h] + 2\lambda_{hH^+H^-}\lambda_{hhH^+H^-} (B_0[p_1^2;h,H^\pm] +B_0[p_2^2;h,H^\pm]) \notag\\
  &+ 2\lambda_{hHH}\lambda_{HHH^+H^-} B_0[q^2;H,H] + \lambda_{HG^+H^-}\lambda_{hHH^+G^-} (B_0[p_1^2;H,G^\pm] +B_0[p_2^2;H,G^\pm]) \notag\\
  &+ 2\lambda_{hAA}\lambda_{AAH^+H^-} B_0[q^2;A,A] + \lambda_{AG^+H^-}\lambda_{hAH^+G^-} (B_0[p_1^2;A,G^\pm] +B_0[p_2^2;A,G^\pm]) \notag\\  
  &+ 2\lambda_{hG^0G^0}\lambda_{G^0G^0H^+H^-} B_0[q^2;G^0,G^0] + 4\lambda_{hH^+H^-}\lambda_{H^+H^-H^+H^-} B_0[q^2;H^\pm,H^\pm]) \notag\\
  &+ \lambda_{hG^+G^-}\lambda_{H^+H^-G^+G^-} B_0[q^2;G^\pm,G^\pm]) \notag\\
  & +2g^3 m_W(B_0[q^2;W,W] -\frac{1}{2}) + c_{2W}^2g_Z^3 m_Z(B_0[q^2;Z,Z] -\frac{1}{2}) \notag\\
  & -\frac{g^4v}{8}C_{VSV}^{\phi\phi\phi}[W,H,W] - \frac{g^4v}{8}C_{VSV}^{\phi\phi\phi}[W,A,W] - \frac{g_Z^4v}{8}c_{2W}^{2}C_{VSV}^{\phi\phi\phi}[Z,H^+,Z] \notag\\
  & +\frac{g^2}{2}\lambda_{hHH}C_{SVS}^{\phi\phi\phi}[H,W,H] + \frac{g^2}{2}\lambda_{hAA}C_{SVS}^{\phi\phi\phi}[A,W,A]
  +\frac{g_Z^2}{4}c_{2W}^2\lambda_{hH^+H^-}C_{SVS}^{\phi\phi\phi}[H^+,Z,H^+]\notag\\
  &+ e^2 \lambda_{hH^+H^-} C_{VSV}^{\phi\phi\phi}[H^+,\gamma,H^+] 
  - \frac{g^2}{4}\lambda_{HH^-G^+}(C_{VSS}^{\phi\phi\phi}[W,H,G^+] + C_{SSV}^{\phi\phi\phi}[G^+,H,W]) \notag\\
  & +i\frac{g^2}{4}\lambda_{AH^-G^+}^{}(C_{VSS}^{\phi\phi\phi}[W,A,G^+] +C_{SSV}^{\phi\phi\phi}[G^+,A,W]) \notag\\
  & -6\lambda_{hhh}\lambda_{hH^+H^-}^2C_0[h,H^+,h] -2\lambda_{hHH}\lambda_{HH^+G^-}\lambda_{HH^-G^+}C_0[H,G^+,H]
  \notag\\
  &-2\lambda_{hAA}\lambda_{AH^+G^-}\lambda_{AH^-G^+}C_0[A,G^+,A]
   - \lambda_{hH^+H^-}^3 C_0[H^+,h,H^+] \notag\\
   &-\lambda_{hG^+G^-}\lambda_{HH^+G^-}\lambda_{HH^-G^+}C_0[G^+,H,G^+]
  -\lambda_{hG^+G^-}\lambda_{AH^+G^-}\lambda_{AH^-G^+}C_0[G^+,H,H^+]. 
\end{align}

\section{$\phi \to H^+H^- \gamma$}\label{app:real emission}
We here give the explicit formula of the real photon emission process for $\phi \to H^+H^-$, which cancels IR divergence of the virtual corrections for the $\phi H^+H^-$ vertex. 
The formula of the decay rate can be expressed as
\begin{align}
  \Gamma[\phi\to H^+H^-\gamma]
  &=\frac{e^2\lambda_{\phi H^+H^-}^2}{16\pi^3m_\phi^{}}\big\{
  -I_1 - I_2 -m_{H^\pm}^2I_{11}^{} -m_{H^\pm}^2I_{22}^{} 
  -(2m_{H^\pm}^2 -m_\phi^2)I_{12}^{}\big\}, \label{eq:real_emission}
\end{align}
where definitions of $I$-functions are given in Appendix D of Ref.~\cite{Denner:1991kt}.

\bibliographystyle{apsrev4-1}
\bibliography{bibhcoup}
\end{document}